\documentclass[11pt,hyper,a4paper]{article}
\usepackage{jheppub}
\usepackage{amsmath,amssymb,amsfonts,amscd,bm, amsthm, slashed, mathrsfs, bm, bbm}
\usepackage{graphicx, wrapfig}
\usepackage{multirow}
\usepackage{verbatim}
\usepackage[title]{appendix}
\usepackage{fancybox}
\usepackage[utf8]{inputenc}
\usepackage{subcaption}
\usepackage[bbgreekl]{mathbbol}

\usepackage{url}
\usepackage{float}

\usepackage[normalem]{ulem}
\usepackage{graphicx}

\usepackage{color}
\usepackage[usenames,dvipsnames,svgnames,table]{xcolor}
\definecolor{darkblue}{cmyk}{0.9,0.9,0,0}
\hypersetup{colorlinks=false, linkcolor=darkblue,linkcolor=darkblue, citecolor=darkblue}

\usepackage{multirow}
\usepackage{verbatim}

\usepackage{bbm}
\usepackage{comment}
\usepackage{enumitem}
\usepackage{float}
\usepackage{empheq}
\usepackage[usenames,dvipsnames,svgnames,table]{xcolor}
\usepackage{enumerate}

\title{Resurgent Analysis for Some 3-manifold Invariants}

\author{Hee-Joong Chung}

\affiliation{Yau Mathematical Sciences Center, Tsinghua University, Beijing 100084, China}

\abstract{
We study resurgence for some 3-manifold invariants when $G_{\mathbb{C}}=SL(2, \mathbb{C})$.
We discuss the case of an infinite family of Seifert manifolds for general roots of unity and the case of the torus knot complement in $S^3$.
Via resurgent analysis, we see that the contribution from the abelian flat connections to the analytically continued Chern-Simons partition function contains the information of all non-abelian flat connections, so it can be regarded as a full partition function of the analytically continued Chern-Simons theory on 3-manifolds $M_3$.
In particular, this directly indicates that the homological block for the torus knot complement in $S^3$ is an analytic continuation of the full $G=SU(2)$ partition function, \textit{i.e.} the colored Jones polynomial.
}

\begin{document}

\maketitle

\section{Introduction}

Many exact results on the partition function and its perturbative expansions are available in Chern-Simons theory, so it provides various examples for resurgent analysis.
A number of interesting aspects of Chern-Simons theory in the context of resurgence have been studied in \cite{Gukov-Marino-Putrov}.
For example, it was argued that all non-abelian flat connections are attached to the abelian flat connections, \textit{i.e.} it is possible to recover the information of all non-abelian flat connections from the perturbative expansions around the abelian flat connections on 3-manifolds $M_3$ via the Stokes phenomena.
This was checked for some Brieskorn spheres, which are integer homology spheres.
Since they are integer homology spheres, the Witten-Reshetikhin-Turaev (WRT) invariant or the Chern-Simons (CS) partition function on them is given by a homological block introduced in \cite{Gukov-Putrov-Vafa, Gukov-Pei-Putrov-Vafa}.

Also, aspects of resurgence in Chern-Simons theory via the modularity of false theta functions have been studied in \cite{Cheng-Chun-Ferrari-Gukov-Harrison} and the contributions from $SL(2,\mathbb{C})$ flat connections to the SU(2) WRT invariant were calculated for some Seifert rational homology spheres.

Contributions from the abelian flat connections to the partition function for an infinite class of Seifert manifolds with a gauge group $G=SU(N)$, $N \geq 2$, have been calculated in \cite{Chung-Seifert}.
In particular, when $G=SU(2)$, it was noted that the integral expression of the contribution from an abelian flat connection, which is obtained after the analytic continuation of the level $K$ in the expression obtained in \cite{Lawrence-Rozansky} with an appropriate choice of the integration contour, is the same as the Borel resummation of the Borel transform of the perturbative expansion around the abelian flat connection under consideration.
Similar discussion for the case of a rational level $K$ has also been discussed in \cite{Chung-rationalk} but without considering resurgence explicitly.

Meanwhile, a new two-variable invariant for the knot complement in $S^3$ has been discussed in \cite{Gukov-Manolescu}, which is denoted $F_\mathcal{K}(x;q)$ for a knot $\mathcal{K}$.
This is obtained from resurgent analysis on the perturbative expansion around the abelian flat connection, which is trivial, where a thorough resurgent analysis for $F_\mathcal{K}(x;q)$ has not been discussed.	\\

In this paper, we consider resurgence for the Chern-Simons partition functions on an infinite family of the Seifert manifolds with an integer and a rational level $K$, and the torus knot complement in $S^3$ when $G_{\mathbb{C}}=SL(2,\mathbb{C})$.
In section 2, we first review and study the case of the Seifert integer homology spheres when the level $K$ is taken to be an integer, which serves as a basic example for the rest of the paper.
Then we consider the case of Seifert rational homology spheres.
As in the case of Seifert integer homology spheres, we see that the contributions from all non-abelian flat connections can be obtained from the contributions from abelian flat connections via the Stokes phenomena.
This indicates that the contributions from abelian flat connections with an analytic continuation of the level $K$ can be interpreted as a full partition function of an analytically continued Chern-Simons theory.
We also calculate transseries parameters for an example when $K$ is taken to be an integer and discuss the case of a complex $K$ in Appendix \ref{app:h2g}.

In section 3, we do a similar analysis when we take the level $K=\frac{r}{s}$ to be a rational number where $r$ and $s$ are coprime.
In this case, the structure of the WRT invariant in terms of homological blocks is similar to the case of an integer $K$ but it involves the sum over certain lifts of abeilan flat connections.
We also see that the contributions from non-abelian flat connections can be recovered from the contributions from abelian flat connections.
The calculation of transseries parameters for an example when $K$ is taken to be a rational number is discussed in Appendix \ref{app:h2g}.

In section 4, we first obtain $F_{\mathcal{K}}(x;q)$ for the torus knot complement in $S^3$ from the integral expression in \cite{Lawrence-Rozansky} after analytic continuations of the level $K$ and the color $R$ with an appropriate choice of the contour, which agrees with the Borel resummation of the Borel transform of the perturbative expansion around the abelian flat connection.
We see via resurgent analysis that $F_{\mathcal{K}}(x;q)$ contains information of contributions from all non-abelian flat connections so it is a full partition function of an analytically continued Chern-Simons theory on $S^3 \backslash \mathcal{K}$, \textit{i.e.} an analytic continuation of the colored Jones polynomial.
In addition, we discuss the case of rational $K$.
Also, we provide brief remarks on the modularity and a limit with a general $x$.


\section{Resurgent analysis for Seifert rational homology spheres with integer level $K$}
\label{sec:integerK}

We consider resurgent analysis of the $G=SU(2)$ WRT invariant on an infinite family of the Seifert rational homology spheres $X(P_1/Q_1, \ldots, P_F/Q_F)$ with $F$ singular fibers where $P_j$ and $Q_j$ are coprime for each $j$ and $P_j$'s are pairwise coprime, when the level $K$ is an integer.

One of observations in \cite{Chung-Seifert} is that after an analytic continuation of the level $K$ and the change of the contour the integral expression of the partition function \cite{Lawrence-Rozansky} of Chern-Simons theory on the Seifert manifolds $X(P_1/Q_1, \ldots, P_F/Q_F)$ agrees with the Borel resummation of the exact Borel transform of the perturbative expansion around the abelian flat connections.
It was also checked for the Seifert integer homology spheres that contributions from the non-abelian flat connections to the partition function calculated in \cite{Lawrence-Rozansky} agree with the residues that are captured when $K$ is taken to be an integer in \cite{Gukov-Marino-Putrov}.\footnote{See also \cite{2018arXiv181105376E} for resurgent analysis of the WRT invariants for Seifert integer homology spheres with arbitrary number of singular fibers.}

For discussion on the case of the Seifert rational homology spheres, $H = \pm |\text{Tor}(H_1(M_3), \mathbb{Z})| \geq 2$, we first consider the case of the Seifert integer homology spheres \cite{Chung-Seifert, Gukov-Marino-Putrov}.
We begin with a known exact result of \cite{Lawrence-Rozansky} for the WRT invariant of the Seifert manifolds $X(P_1/Q_1, \ldots, P_F/Q_F)$, and see that contributions from all non-abelian flat connections can be recovered from the homological block via the Stokes phenomena.


\subsection{Resurgent analysis for $H =1$}
\label{ssec:resurg-h1}

For the Seifert manifolds $X(P_1,Q_1, \ldots, P_F, Q_F)$ the full exact $G=SU(2)$ partition function \cite{Lawrence-Rozansky} is given by
\begin{align}
\begin{split}
& \sum_{t=0}^{H-1} \int_{C_0'} f(y) e^{K g_t(y)} dy	- 2 \pi i \sum_{m=1}^{2P-1} \text{Res} \bigg( \frac{f(y) e^{K g_0(y)}}{1-e^{-2Ky}}, y=\pi i m \bigg) 
\end{split}	\label{wrt-su2-orig1}	\\
\begin{split}
&= \int_{C_0'} f(y) e^{K g_0(y)} dy + \sum_{t=1}^{H-1} \int_{C_t'} f(y) e^{K g_t(y)} dy 	\\
& \quad - 2 \pi i \sum_{m=1}^{2P-1} \text{Res} \bigg( \frac{f(y) e^{K g_0(y)}}{1-e^{-2Ky}}, y=\pi i m \bigg) - 2 \pi i \sum_{t=1}^{H-1} \sum_{m=1}^{\lfloor \frac{2Pt}{H} \rfloor} \text{Res} \Big( f(y) e^{K g_t(y)}, y = -\pi i m \Big)
\end{split}	\label{wrt-su2-orig2}
\end{align}
where 
\begin{align}
P = \prod_{j=1}^F P_j	\,	,	\qquad
H = \pm |\text{Tor } H_1(M_3,\mathbb{Z}) | = P \prod_{j=1}^{F} \frac{Q_j}{P_j}, 	
\end{align}
and
\begin{align}
f(y) = \frac{\prod_{j=1}^F (e^{\frac{1}{P_j}y} - e^{-\frac{1}{P_j}y})}{(e^{y} - e^{-y})^{F-2}}	\,	,	\qquad
g_t(y) = -\frac{1}{2\pi i}\frac{H}{P}y^2 - 2ty	\,	.
\end{align}
The Gaussian integral parts in \eqref{wrt-su2-orig2} are the contributions from the abelian flat connections. 
A Weyl orbit of $t$ corresponds to the abelian flat connection of $G=SU(2)$ whose holonomy for the central element in the fundamental group $\pi_1(M_3)$ is $\text{diag} (e^{2\pi i \frac{P}{H}t},e^{-2\pi i \frac{P}{H}t})$ where the Weyl action is given by $t \leftrightarrow -t$ mod $H$, and  $t=0$ corresponds to the trivial flat connection.

The residue parts are the contributions from the non-abelian flat connections.
There are two types of residues at the second line of \eqref{wrt-su2-orig2}.
The second type of residues in \eqref{wrt-su2-orig2} arises due to a parallel shift of the integral contour $C_0'$, which is a line from $-(1+i)\infty$, the origin, and to $(1+i)\infty$ in the $y$-plane, to the contour $C_t'$ that passes the saddle points, $y=-2\pi i \frac{P}{H}t$.	\\

We consider the case of $F=3$ for concreteness.
The case of more singular fibers can be done similarly as in the case of $F=3$.
When $F=3$, the first type of residues in \eqref{wrt-su2-orig2} is given by
\begin{align}
-2 \pi i \sum_{m=1}^{2P-1} 2 i (-1)^m H e^{-\frac{\pi i}{2} K \frac{H}{P} m^2} \Big( \frac{m}{P} - \frac{1}{H} + \frac{i}{H K} \sum_{j=1}^3 \frac{1}{P_j} \cot \frac{\pi m}{P_j} \Big) \prod_{j=1}^3 \sin \frac{\pi m}{P_j}	\,	.	\label{resH}
\end{align}
The term that contains the factor $\frac{i}{H K} \sum_{j=1}^3 \frac{1}{P_j} \cot \frac{\pi m}{P_j}$ vanishes.
$\prod_{j=1}^3 \sin \frac{\pi m}{P_j}$ is nonzero when $m$ is coprime to all $P_j$'s.
Therefore, if $a$ is coprime to all $P_j$'s, then so is $2P-a$.
Hence, the term containing the factor $\frac{1}{H} \prod_{j=1}^3 \sin \frac{\pi m}{P_j}$ in \eqref{resH} vanishes because if $m=a$ gives a nonzero contribution, then $2P-a$ also gives the same value but with an opposite sign.
Thus, the first type of residues in \eqref{wrt-su2-orig2} becomes \cite{Chung-Seifert}
\begin{align}
\sum_{m=1}^{2P-1} 4 \pi (-1)^m H e^{-\frac{\pi i}{2} K \frac{H}{P} m^2} \frac{m}{P} \prod_{j=1}^3 \sin \frac{\pi m}{P_j}	\,	.	\label{su2-h-res}
\end{align}
In particular, when $H=1$ and $F=3$, \eqref{wrt-su2-orig2} is simplified to 
\begin{align}
\begin{split}
\int_{C_0'} f(y) e^{K g_0(y)} dy 
- 2 \pi i \sum_{m=1}^{2P-1} \text{Res} \bigg( \frac{f(y) e^{K g_0(y)}}{1-e^{-2Ky}}, y= \pi i m \bigg) 
\end{split}	\label{wrt-su2-orig-h1}
\end{align}
and from \eqref{su2-h-res} the residue part is given by
\begin{align}
\sum_{m=1}^{2P-1}  4 \pi (-1)^m e^{-\frac{\pi i}{2} K \frac{1}{P} m^2} \frac{m}{P} \prod_{j=1}^3 \sin \frac{\pi m}{P_j}	\,	.	\label{su2-h1-res}
\end{align}


\subsubsection*{Resurgent analysis}

We denote a connected component of the $SL(2,\mathbb{C})$ flat connection as 
\begin{align}
\alpha \in \pi_0 (\mathcal{M}_{\text{flat}}(M_3, SL(2,\mathbb{C})))
\end{align} 
and its lift as 
\begin{align}
\bbalpha = (\alpha, S_{\bbalpha}) \in \pi_0 ( \mathcal{M}_{\text{flat}}(M_3,SL(2,\mathbb{C}))) \times \mathbb{Z}	\,	,
\end{align}
which is a critical point of an analytically continued Chern-Simons theory.
Here, $S_{\bbalpha} \in CS(\alpha) + \mathbb{Z}$ denotes the value of a Chern-Simons invariant of $\bbalpha$ in the universal cover, while $CS(\alpha)$ is a Chern-Simons invariant of $\alpha$, which is defined in mod 1.
A lift of $\alpha$ to $\bbalpha$ is determined by $S_{\bbalpha}$.
Taking $K=|K|e^{i\theta}$, a path integral over a Lefschetz thimble for $\bbalpha$ with a given $\theta$ is denoted by $I_{\bbalpha}(K):=I_{\bbalpha}(|K|,\theta)$.

It is expected that the partition function of an analytically continued Chern-Simons theory takes a form of 
\begin{align}
Z_{M_3}(K) = \sum_{a \in \pi_0 (\mathcal{M}_{\text{flat}}^{\text{ab}}(M_3,SU(2)))} I_{\mathbb{a}}(K)		\label{csab}
\end{align}
where 
\begin{align}
I_{\mathbb{a}}(K) = e^{2 \pi i K S_{\mathbb{a}}} Z_{a}(K)	\label{ia}
\end{align} 
with a chosen lift from $a$ to $\mathbb{a}$ \cite{Gukov-Marino-Putrov}.

From an analytic continuation of the level $K$ in the Gaussian integral part of \eqref{wrt-su2-orig2} with a choice of the integration contour $\gamma$ that is parallel to the imaginary axis of the complex $y$-plane, the contributions $Z_t$ from the abelian flat connections to the partition function of the analytically continued Chern-Simons theory can be obtained \cite{Chung-Seifert}.
These $Z_t$'s that are labelled by abelian flat connections can be further decomposed into the homological blocks with an $SL(2,\mathbb{Z})$ $S$-transform.
We would like to see that contributions from all non-abelian flat connections can be recovered from $I_{\mathbb{t}}$ via the Stokes phenomena, which implies that \eqref{csab} indeed can be regarded as a full partition function.	\\

For the integer homology sphere, there is one abelian flat connection, which is trivial, so $Z_{0}$ is given by a homological block.
When the number of singular fibers is 3, $F=3$, the homological block is given by a linear combination with integer coefficients of the false theta function $\widetilde{\Psi}^{(a)}_P(q)$,
\begin{align}
\widetilde{\Psi}^{(a)}_P(q) = \sum_{n = 1}^{\infty} \psi_{2P}^{(a)}(n) q^{\frac{n^2}{4P}}	\label{ft32}
\end{align}
where
\begin{align}
\psi_{2P}^{(a)}(n) = 
\begin{cases} 
\pm 1	&	n = \pm a \, \text{ mod } 2P	\\ 
0		&	\text{otherwise} 
\end{cases}	\,	.
\end{align}
In particular, when $H=1$, $I_{0}$ for the Seifert manifolds $X(P_1/Q_1, P_2/Q_2, P_3/Q_3)$ was calculated in \cite{Chung-Seifert},
\begin{align}
Z_{M_3}(K)
= \frac{B}{2i} q^{-\phi_3/4} \Big( \frac{2i}{K} P \Big)^{1/2} 
\times 
\begin{cases}
\sum_{j=0}^3 \widetilde{\Psi}^{(R_j)}_P(q)	&	\text{when }	\sum_{j=1}^{3} \frac{1}{P_j} < 1		\\
2q^{\frac{P}{4} \big( \sum_{j=1}^{3}\frac{1}{P_j}-1 \big)^2} + \sum_{j=0}^3 \widetilde{\Psi}^{(R_j)}_P(q)	&	\text{when }	\sum_{j=1}^{3} \frac{1}{P_j} > 1		
\end{cases}	\label{hbl-h1}
\end{align}
where
\begin{align}
B	&=	-\frac{\text{sign}P}{4 \sqrt{|P|}} e^{\frac{3}{4}\pi i \, \text{sign} \, \left( \frac{H}{P} \right)}	\,	,	\\
\phi_F	&=	3 \, \text{sign}\left( \frac{H}{P} \right) + \sum_{j=1}^{F} \left( 12 s(Q_j, P_j) - \frac{Q_j}{P_j} \right)		\,	,
\end{align}
and $s(Q,P)$ is the Dedekind sum
\begin{align}
s(Q,P) = \frac{1}{4P} \sum_{l=1}^{P-1} \cot \Big( \frac{\pi l}{P} \Big) \cot \Big( \frac{\pi Q l}{P} \Big)	\,	
\end{align}
for $P>0$, which satisfies $s(-Q,P)=-s(Q,P)$.
$R_j$, $j=0,1,2,3$, in \eqref{hbl-h1} are given by
\begin{equation}
\begin{aligned}
&R_0 = P(1-(1/P_1 + 1/P_2 + 1/P_3)),	&&	\quad	R_1 = P(1-(1/P_1 -1/P_2 -1/P_3)), \\
&R_2 = P(1 - (-1/P_1 + 1/P_2 - 1/P_3)),	&&	\quad	R_3 = P(1 - (-1/P_1 - 1/P_2 + 1/P_3))	\,	.	\label{rj}
\end{aligned}
\end{equation}
In \eqref{hbl-h1}, $\sum_{j=1}^3 \frac{1}{P_j}>1$ is satisfied only when $(P_1,P_2,P_3)=(2,3,5)$.
Also, we see that the Chern-Simons invariant for the abelian flat connection is zero, and we chose a lift such that $S_{\mathbb{0}}=CS(0)=0$.

\begin{figure}
\begin{subfigure}{0.3\textwidth}
\centering
\includegraphics[width=40mm]{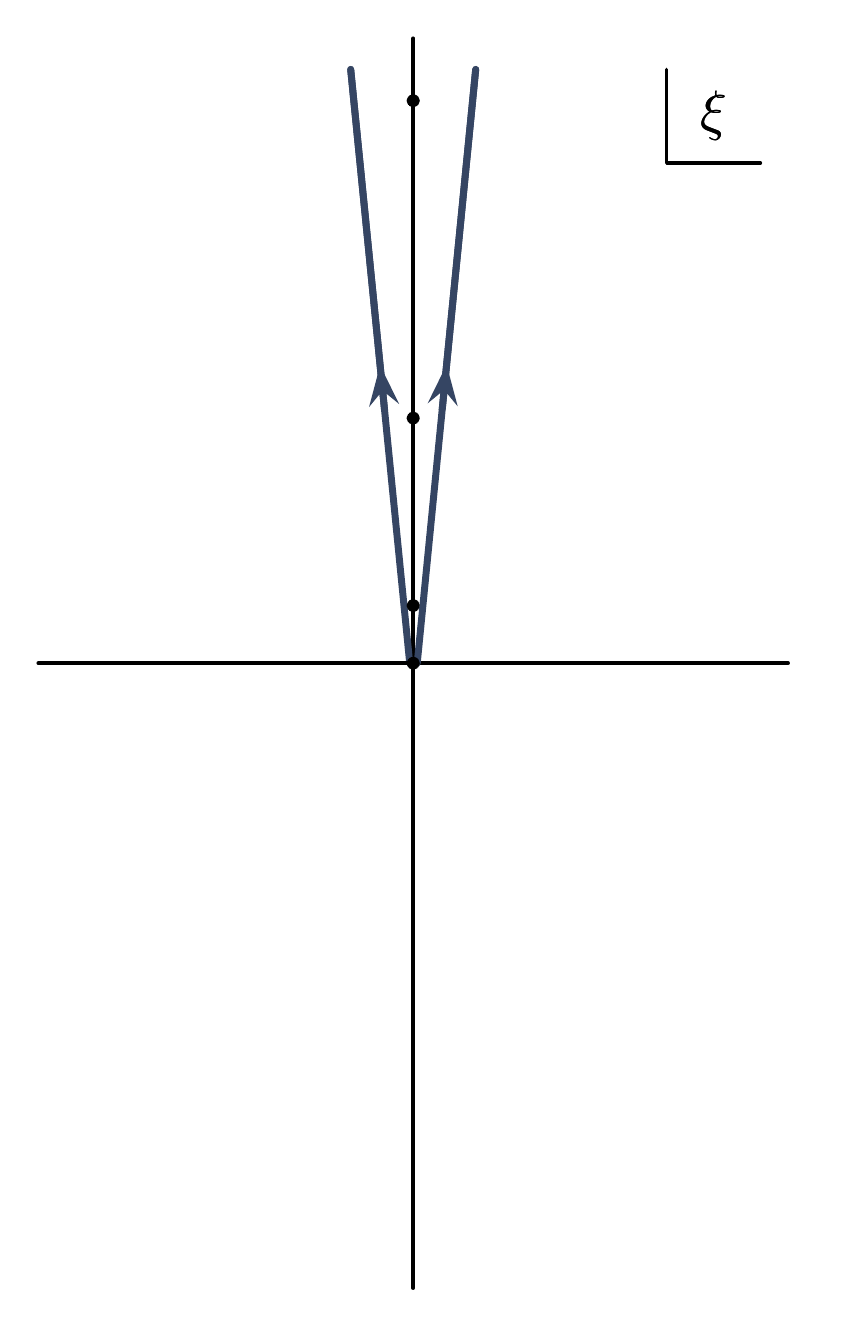} 
\caption{}	\label{resum1}
\end{subfigure}
\begin{subfigure}{0.3\textwidth}
\centering
\includegraphics[width=40mm]{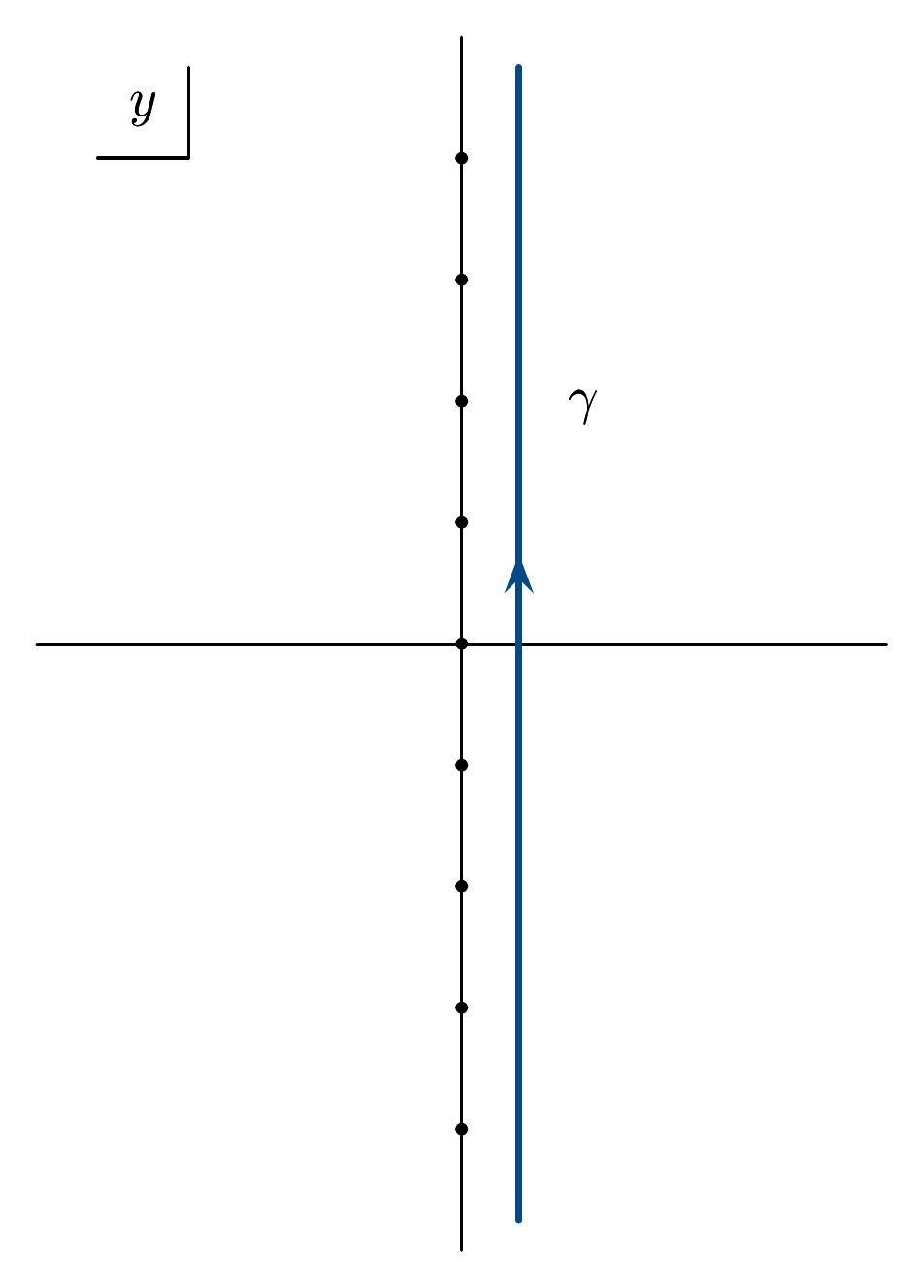} 
\caption{}	\label{resum2}
\end{subfigure}
\begin{subfigure}{0.4\textwidth}
\centering
\includegraphics[width=60mm]{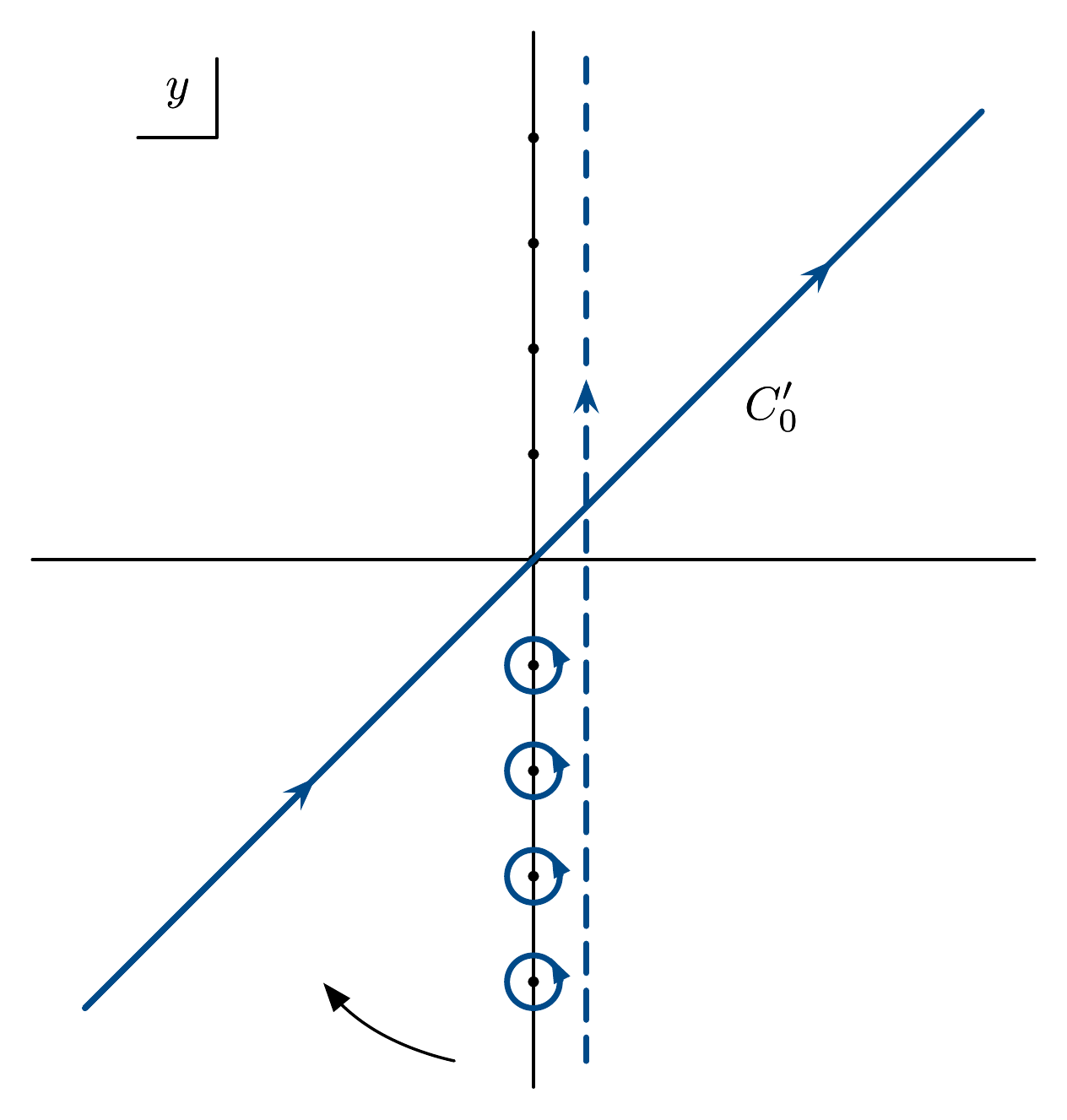} 
\caption{}	\label{resum3}
\end{subfigure}
\caption{The integration contours}
\label{fig:Bresum}
\end{figure}

The integral part of \eqref{wrt-su2-orig2} with the contour $\gamma$ that calculates $I_{0}$ agrees with the Borel resummation after the change of integration variable.
For $\widetilde{\Psi}^{(a)}_P(q)$, the Borel resummation is given by the average of the Borel sums \cite{Gukov-Marino-Putrov},
\begin{align}
Z(q) = \frac{1}{2} \Big( \mathcal{S}_{\frac{\pi}{2} - \delta} Z_{\text{pert}} + \mathcal{S}_{\frac{\pi}{2} + \delta} Z_{\text{pert}} \Big)	\,		\label{resurg-H1}
\end{align}
where $\frac{\pi}{2} \pm \delta$ denotes the integration contours (\textit{c.f.} the contours in Figure \ref{resum1}) in the Borel plane, more specifically,
\begin{align}
\frac{1}{\sqrt{K}} \widetilde{\Psi}^{(a)}_P(q) = \frac{1}{2} \bigg( \int_{i e^{+ i \delta} \mathbb{R}_+}+ \int_{i e^{- i \delta} \mathbb{R}_+} \bigg) \frac{d\xi}{\sqrt{\pi \xi}} \frac{\sinh(P-a) \big( \frac{2 \pi i \xi}{P} \big)^{1/2}}{\sinh P \big( \frac{2 \pi i \xi}{P} \big)^{1/2}} e^{-K \xi}	\,	.	\label{resurg-psi}
\end{align}
With a change of a variable $y= ( 2 \pi i P \xi )^{1/2}$, the Borel resummation \eqref{resurg-psi} is written as
\begin{align}
\frac{1}{P} \int_{\gamma} dy \, e^{-\frac{1}{2\pi i} \frac{1}{P} K y^2} \frac{ e^{(P-a)y/P} - e^{-(P-a)y/P}}{(e^{y} - e^{-y})}  	\label{resurg-CS}
\end{align}
where the contour in the Borel plane in Figure \ref{resum1} becomes the contour $\gamma$ in Figure \ref{resum2}.
Therefore, we can see that with $y= ( 2 \pi i P \xi )^{1/2}$, the Borel resummation \eqref{resurg-H1} for the Seifert manifolds with three singular fibers is the same as the integral part of \eqref{wrt-su2-orig-h1} with the integration contour $\gamma$ and $\text{Im }K < 0$,
\begin{align}
\frac{1}{P} \int_{\gamma} dy \, e^{-\frac{1}{2\pi i} \frac{1}{P} K y^2} \frac{\prod_{j=1}^3 (e^{y/P_j} - e^{-y/ P_j})}{(e^{y} - e^{-y})}	\,	.	\label{resurg-CS-f3}
\end{align}
It is more convenient to consider the Borel resummation as in the expression \eqref{resurg-CS-f3} rather than as in the standard expression \eqref{resurg-psi} in the Borel plane, so we consider the former type of expression in the rest of the paper.	\\

If $K \in \mathbb{Z}_{+}$, the contour should be changed so that the integral is convergent and such an integration contour is given by a line that passes from $-(1+i)\infty$, the origin, and to $(1+i) \infty$ in the $y$-plane as in Figure \ref{resum3}.
Due to such a move of the contour, poles at the negative imaginary axis of the $y$-plane are picked up and their contributions to the residues are given by
\begin{align}
2 \pi i \sum_{m=1}^{\infty} \text{Res} \bigg[ e^{-\frac{1}{2\pi i} \frac{1}{P} K y^2} \frac{\prod_{j=1}^3 (e^{\frac{y}{P_j}} - e^{-\frac{y}{P_j}})}{(e^{y} - e^{-y})}, y=-m \pi i \bigg]  
= -8 \pi \sum_{m=1}^{\infty} (-1)^m e^{-\frac{\pi i}{2}K\frac{1}{P} m^2} \prod_{j=1}^{3} \sin \frac{m \pi}{P_j}	\,	.	\label{res-1}
\end{align}
So $m$'s that are not multiples of $P_j$'s give non-zero contributions in \eqref{res-1}.
For example, when $(P_1,P_2,P_3)=(2,3,5)$, they include $m=1,11,19,29,31,41,49,59$ mod $60$ and $m=7,13,17,23,37,43,47,53$ mod $60$.
Each set of poles corresponds to non-abelian flat connections whose Chern-Simons invariants are $-\frac{1}{120}$ and $-\frac{49}{120}$, mod 1, respectively.

The infinite sum in \eqref{res-1} can be regularized.
For example, consider a periodic function $C_{p}$ with a period $p$, $C_p(n+p) = C_p(n)$, which has 0 mean value $\sum_{n=0}^{p-1}C_p(n)=0$.
Then it is known \cite{Lawrence-Zagier} that an $L$-function $L(s,C_p) = \sum_{n=1}^{\infty} C_p(n) n^{-s}$ associated to $C_p$ where $\text{Re } s > 1$ is asymptotically
\begin{align}
\sum_{n=0}^{\infty} C_p(n) e^{-n^2 w} \simeq \sum_{n=0}^{\infty} L(-2n, C_p) \frac{(-w)^n}{n!}	\label{l-ftn}
\end{align}
as $w$ approaches to $0$ from the upper half plane, $w \searrow 0$.
Here, $L(-r, C_p)$ is given by
\begin{align}
L(-r, C_p) = -\frac{p^r}{r+1} \sum_{n=1}^{p} C_p(n) B_{r+1} \Big( \frac{n}{p} \Big)
\end{align}
where $B_{j}(x)$ is the Bernoulli polynomial whose generating function is $\frac{u e^{ux}}{e^u-1} = \sum_{m=0}^{\infty} B_m(x) \frac{u^m}{m!}$.
We consider $Y(w) = \sum_{n=0}^{\infty} \psi_{2P}(n) e^{-\frac{\pi i}{2} \frac{1}{P} K n^2} e^{-n^2 w}$. 
Taking $C_{2P}(n) = \psi_{2P}^{(a)}(n) e^{-\frac{\pi i}{2} \frac{1}{P} K n^2}$, $C_{2P}(n)$ is a periodic function with a period $2P$ and its mean value is zero because $C_{2P}(2P-n) = -C_{2P}(n)$.
Then from \eqref{l-ftn}, $Y(w) \simeq \sum_{n=0}^{\infty} L(-2n, C_{2P}) \frac{(-w)^n}{n!}$ as $w \searrow 0$, so
\begin{align}
Y(0) \simeq L(0,C_{2P})= - \sum_{n=1}^{2P} C_{2P}(n) B_1 \Big( \frac{n}{2P} \Big) = \frac{1}{2} \sum_{n=1}^{2P} \psi_{2P}^{(a)} e^{-\frac{\pi i}{2} \frac{1}{P} K n^2} \Big( 1- \frac{n}{P} \Big)
\end{align}
Therefore, 
\begin{align}
\sum_{n=0}^{\infty} \psi^{(a)}_{2P}(n) e^{-\frac{\pi i}{2} \frac{1}{P} K n^2} = \frac{1}{2} \sum_{n=1}^{2P} \psi_{2P}^{(a)}(n) e^{-\frac{\pi i}{2} \frac{1}{P} K n^2} \big( 1- \frac{n}{P} \big)	\,	.
\label{reg}
\end{align}
Or simply we may consider a limit $y \searrow 0$ on $\frac{\sinh (P-a) y}{\sinh Py} = \sum_{n=1}^{\infty} \psi_{2P}^{(a)}(n) e^{-ny}$, which gives a regularization $\sum_{n=1}^{\infty} \psi_{2P}^{(a)}(n) \simeq 1-\frac{a}{P}$.

For $m=a$ that is not a multiple of $P_j$'s, the contribution from $m=a$ in \eqref{res-1} has an opposite sign of the contribution from $m=2P-a$ in \eqref{res-1}.
Thus, \eqref{res-1} can be regularized to
\begin{align}
-8 \pi \sum_{m= \pm a \text{ mod } 2P}^{\infty}  (-1)^m e^{-\frac{\pi i}{2}K\frac{1}{P} m^2} \prod_{j=1}^{3} \sin \frac{m \pi}{P_j}
= -\sum_{m=a, \, 2P-a}
4 \pi (-1)^m e^{-\frac{\pi i}{2}K\frac{1}{P} m^2} \Big(1- \frac{m}{P} \Big) \prod_{j=1}^{3} \sin \frac{m \pi}{P_j}	\,	,	\label{res-2}
\end{align}
which is further simplified to
\begin{align}
\sum_{m=a, 2P-a} 
4 \pi (-1)^m e^{-\frac{\pi i}{2}K\frac{1}{P} m^2} \frac{m}{P} \prod_{j=1}^{3} \sin \frac{m \pi}{P_j}	\,	.
\end{align}
Therefore, we have
\begin{align}
-8 \pi \sum_{m= 1}^{\infty}  (-1)^m e^{-\frac{\pi i}{2}K\frac{1}{P} m^2} \prod_{j=1}^{3} \sin \frac{m \pi}{P_j}
= \sum_{m=1}^{2P}
4 \pi (-1)^m e^{-\frac{\pi i}{2}K\frac{1}{P} m^2} \frac{m}{P} \prod_{j=1}^{3} \sin \frac{m \pi}{P_j}	\,	,	\label{res-3}
\end{align}
which agrees with \eqref{su2-h1-res}.

Transseries parameters for some examples of $H=1$ have been calculated in \cite{Gukov-Marino-Putrov, Chun:2017dbf}, so for the case of $H=1$ we don't provide further examples. 
Instead, we calculate transseries parameters when $H \geq 2$ in section \ref{ssec:rsgh2}.


\subsection*{A remark on modularity from resurgent analysis}

As a simple consistency check, we also discuss a relation between the modularity and the resurgent analysis.
It is known that the false theta function $\widetilde{\Psi}_{P}^{(a)}(q)$ satisfies a nearly modular property \cite{MR2048564, Lawrence-Zagier}
\begin{align}
\widetilde{\Psi}_{P}^{(a)}(q=e^{2\pi i \frac{1}{K}}) = -\sqrt{\frac{K}{i}} \sum_{b=1}^{P-1} M_{ab} \widetilde{\Psi}_{P}^{(b)}(e^{-2\pi i K}) + \sum_{n=0}^{\infty} \frac{c_n}{K^n} \Big( \frac{\pi i}{2P} \Big)^n
\end{align}
where $M_{ab} = \sqrt{\frac{2}{P}} \sin \frac{\pi a b}{P}$ and $c_n$ is such that
\begin{align}
\frac{\sinh (P-a) z}{ \sinh Pz} = \sum_{n=0}^{\infty} \frac{n!}{(2n)!} c_n z^{2n}	\,	.
\end{align}
When $K \in \mathbb{Z}$, $\widetilde{\Psi}_{P}^{(a)}(e^{-2\pi i K}) = \big(1-\frac{a}{P} \big) e^{-\frac{\pi i}{2}\frac{1}{P} K a^2}$, so 
\begin{align}
\widetilde{\Psi}_{P}^{(a)}(q=e^{2\pi i \frac{1}{K}}) = - \sqrt{\frac{K}{i}} \sum_{b=1}^{P-1} \sqrt{\frac{2}{P}} \sin \frac{\pi a b}{P} \Big(1-\frac{b}{P} \Big) e^{-\frac{\pi i}{2}\frac{1}{P} K b^2} + \sum_{n=0}^{\infty} \frac{c_n}{K^n} \Big( \frac{\pi i}{2P} \Big)^n	\,	.	\label{mod-H1}
\end{align}
The first and the second term in \eqref{mod-H1}, which are the non-perturbative and the perturbative part, correspond to the contributions from the non-abelian and the abelian flat connection, respectively.	
\eqref{mod-H1} can be derived from \eqref{resurg-psi} when $K$ is taken to be an integer.	\\

For the Seifert manifolds with $H=1$ and $F=3$, by using
\begin{align}
\prod_{j=1}^3 \sin \frac{\pi m}{P_j} = \frac{1}{4} (-1)^m \sum_{j=0}^{3} \sin \frac{\pi R_j m}{P}	\,		\label{sine-id}
\end{align}
where $R_j$'s are given in \eqref{rj}, the contribution \eqref{res-3} from non-abelian flat connections when $K \in \mathbb{Z}_{+}$ can be expressed as
\begin{align}
- \sum_{j=0}^{3} \sqrt{\frac{K}{i}} \sum_{m=1}^{P} \sqrt{\frac{2}{P}}  \sin \frac{\pi R_j m}{P} e^{-\frac{\pi i}{2}K\frac{1}{P} m^2} \Big( 1 -\frac{m}{P} \Big) 	\,	.	\label{nab-mod-h1}
\end{align}
Since $Z_{M_3}(K) \simeq \sum_{j=0}^3 \widetilde{\Psi}^{(R_j)}_P(q) \big|_{q \searrow e^{\frac{2\pi i}{K}}}$ when $F=3$ and $H=1$, \eqref{nab-mod-h1} agrees with the non-perturbative part of the WRT invariant calculated from \eqref{mod-H1} by using the modularity.


\subsection{Resurgent analysis for $H \geq 2$}
\label{ssec:rsgh2}

When $H \geq 2$, we would also like to see that the contributions from the non-abelian flat connections can be recovered from the contributions from the abelian flat connections.
In addition, we calculate transseries parameters for an example when $K$ is taken to be an integer.	\\

As discussed in section \ref{ssec:resurg-h1}, the residue $- 2 \pi i \sum_{m=1}^{2P-1} \text{Res} \Big( \frac{f(y) e^{K g_0(y)}}{1-e^{-2Ky}}, y=\pi i m \Big)$ in the full $SU(2)$ WRT invariant \eqref{wrt-su2-orig2} is given by
\begin{align}
4 \pi H (-1)^m e^{-\frac{\pi i}{2} K \frac{H}{P} m^2} \frac{m}{P} \prod_{j=1}^3 \sin \frac{\pi m}{P_j}	\,	.
\end{align}
The other type of residue in \eqref{wrt-su2-orig2} is given by
\begin{align}
- 2 \pi i \sum_{t=1}^{H-1} \sum_{m=1}^{\lfloor \frac{2Pt}{H} \rfloor} \text{Res} \Big( f(y) e^{K g_t(y)}, y = -\pi i m \Big) 
= - \sum_{t=1}^{H-1} \sum_{m=1}^{\lfloor \frac{2Pt}{H} \rfloor} \pi i (-1)^m e^{-\frac{\pi i}{2}K \frac{H}{P} m^2} \prod_{j=1}^3 \big( e^{m \pi i / P_j} - e^{-m \pi i / P_j} \big) 
\end{align}
Therefore, the total residue, \textit{i.e.} the total contribution from the non-abelian flat connections to the WRT invariant, is
\begin{align}
\sum_{m=1}^{2P} 4 \pi H (-1)^m e^{-\frac{\pi i}{2} K \frac{H}{P} m^2} \frac{m}{P} \prod_{j=1}^3 \sin \frac{\pi m}{P_j}  
+ 8 \pi (-1)^m \sum_{t=1}^{H-1} \sum_{m=1}^{\lfloor \frac{2Pt}{H} \rfloor} \prod_{j=1}^3 \sin \frac{\pi m}{P_j} \, e^{-\frac{\pi i}{2}K \frac{H}{P} m^2}		\,	.	\label{wrt-res-h}
\end{align}


\subsubsection*{Resurgent analysis}

The Gaussian integral part of \eqref{wrt-su2-orig2} for $K \in \mathbb{Z}_+$ can be written as
\begin{align}
\sum_{t=0}^{H-1} e^{2 \pi i K \frac{P}{H}t^2} \int_{C_t'} dy \, e^{-\frac{K}{2\pi i} \frac{H}{P} \big( y + 2 \pi i \frac{P}{H} t \big)^2} \frac{\prod_{j=1}^F (e^{\frac{1}{P_j}y} - e^{-\frac{1}{P_j}y})}{e^{y} - e^{-y}}	\,	.	\label{int-genh}
\end{align}
From \eqref{int-genh}, it is possible to express the WRT invariant in terms of homological blocks
\begin{align}
\begin{split}
\text{WRT}_{M_3}(K) 
=& \frac{B}{2i} q^{-\phi_3/4} \bigg( \frac{2i}{K} \frac{P}{H} \bigg)^{1/2} 
\Bigg[ \sum_{s=0}^{3} \Big( \sum_{h=0}^{\frac{H-1}{2}} \widetilde{\Psi}^{(2hP+R_s)}_{HP}(q) - \sum_{h=0}^{\frac{H-1}{2}-1} \widetilde{\Psi}^{(2(h+1)P-R_s)}_{HP}(q) \Big) 		\\
&   + \sum_{s=0}^{3} \sum_{t=1}^{\frac{H-1}{2}} e^{2 \pi i K \frac{P}{H} t^2} 
\bigg( \sum_{h=0}^{\frac{H-1}{2}} (e^{-2\pi i \frac{t}{H} (2hP+R_s)} + e^{2\pi i \frac{t}{H} (2hP+R_s)}) \widetilde{\Psi}^{(2hP+R_s)}_{HP}(q) 	\\ 
&\hspace{25mm}	- \sum_{h=0}^{\frac{H-1}{2}-1} (e^{-2\pi i \frac{t}{H} (2(h+1)P-R_s)} + e^{2\pi i \frac{t}{H} (2(h+1)P-R_s)}) \widetilde{\Psi}^{(2(h+1)P-R_s)}_{HP}(q) \bigg) 	\Bigg]	\,	\Bigg|_{q \searrow e^{\frac{2\pi i}{K}}}	\,	.
\end{split}
\label{3f-odd}
\end{align}
when $H$ is odd, and
\begin{align}
\begin{split}
\text{WRT}_{M_3}(K)
=& \frac{B}{2i} q^{-\phi_3/4} \bigg( \frac{2i}{K} \frac{P}{H} \bigg)^{1/2} 
\Bigg[ \sum_{s=0}^{3} \sum_{h=0}^{\frac{H}{2}-1} (\widetilde{\Psi}^{(2hP+R_s)}_{HP} - \widetilde{\Psi}^{(2(h+1)P-R_s)}_{HP} )		\\
& \hspace{-10mm}  + \sum_{s=0}^{3} \sum_{h=0}^{\frac{H}{2}-1} \bigg( \sum_{t=1}^{\frac{H-2}{2}} e^{2 \pi i K \frac{P}{H} t^2} \Big( (e^{-2\pi i \frac{t}{H} (2hP+R_s)} + e^{2\pi i \frac{t}{H} (2hP+R_s)}) \widetilde{\Psi}^{(2hP+R_s)}_{HP} 	\\ 
&\hspace{37mm}	-(e^{-2\pi i \frac{t}{H} (2(h+1)P-R_s)} + e^{2\pi i \frac{t}{H} (2(h+1)P-R_s)}) \widetilde{\Psi}^{(2(h+1)P-R_s)}_{HP} \Big) 	\bigg)	\\
&\hspace{15mm}	+e^{\frac{\pi i}{2} K PH} \Big( e^{\pi i (2hP+R_s)} \widetilde{\Psi}^{(2hP+R_s)}_{HP} - e^{\pi i (2(h+1)P-R_s)} \widetilde{\Psi}^{(2(h+1)P-R_s)}_{HP} \Big)	
\Bigg]	\,	\Bigg|_{q \searrow e^{\frac{2\pi i}{K}}}	\,	.
\end{split}
\label{3f-even}
\end{align}
when $H$ is even \cite{Chung-Seifert}.
Thus, for an odd $H$,
\begin{align}
Z_0(q) &= \sum_{s=0}^{3} \Big( \sum_{h=0}^{\frac{H-1}{2}} \widetilde{\Psi}^{(2hP+R_s)}_{HP}(q) - \sum_{h=0}^{\frac{H-1}{2}-1} \widetilde{\Psi}^{(2(h+1)P-R_s)}_{HP}(q) \Big)	\label{z0} 	\\
\begin{split}
Z_a(q) &= \sum_{s=0}^{3} \bigg( \sum_{h=0}^{\frac{H-1}{2}} (e^{-2\pi i \frac{a}{H} (2hP+R_s)} + e^{2\pi i \frac{a}{H} (2hP+R_s)}) \widetilde{\Psi}^{(2hP+R_s)}_{HP}(q) 	\\ 
&\hspace{12mm}	- \sum_{h=0}^{\frac{H-1}{2}-1} (e^{-2\pi i \frac{a}{H} (2(h+1)P-R_s)} + e^{2\pi i \frac{a}{H} (2(h+1)P-R_s)}) \widetilde{\Psi}^{(2(h+1)P-R_s)}_{HP}(q) \bigg)
\end{split}	\label{zt}
\end{align}
where $a=1, \ldots, \frac{H-1}{2}$ denotes a Weyl orbit $W_a$, which contains $\{a,H-a\}$. 
When $H$ is even, we have
\begin{align}
Z_0(q) &= \sum_{s=0}^{3} \sum_{h=0}^{\frac{H}{2}-1} (\widetilde{\Psi}^{(2hP+R_s)}_{HP} - \widetilde{\Psi}^{(2(h+1)P-R_s)}_{HP} )	\label{z0e} 	\\
Z_{\frac{H}{2}}(q) &= \sum_{s=0}^{3} \sum_{h=0}^{\frac{H}{2}-1} \Big( e^{\pi i (2hP+R_s)} \widetilde{\Psi}^{(2hP+R_s)}_{HP} - e^{\pi i (2(h+1)P-R_s)} \widetilde{\Psi}^{(2(h+1)P-R_s)}_{HP} \Big)	\label{zhalfe} 	\\
\begin{split}
Z_a(q) &= \sum_{s=0}^{3} \sum_{h=0}^{\frac{H}{2}-1} \sum_{t=1}^{\frac{H-2}{2}} \Big( (e^{-2\pi i \frac{t}{H} (2hP+R_s)} + e^{2\pi i \frac{t}{H} (2hP+R_s)}) \widetilde{\Psi}^{(2hP+R_s)}_{HP} 	\\ 
&\hspace{25mm}	-(e^{-2\pi i \frac{t}{H} (2(h+1)P-R_s)} + e^{2\pi i \frac{t}{H} (2(h+1)P-R_s)}) \widetilde{\Psi}^{(2(h+1)P-R_s)}_{HP} \Big) 
\end{split}	\label{zte}
\end{align}
where $a=1, \ldots, \frac{H}{2}-1$.
Below, we consider the case of an odd $H$ for convenience.
The case of even $H$ can be done similarly.	\\

As in the case of $H=1$, we can obtain the Borel resummation of the Borel transform of the perturbative expansions around the abelian flat connection $t$.
More explicitly, when $t \neq 0$, from \eqref{ft32}, \eqref{zt} is expressed as
\begin{align}
\begin{split}
\hspace{-2mm}\sum_{s=0}^{3} \bigg( & \sum_{h=0}^{\frac{H-1}{2}} (e^{-2\pi i \frac{a}{H} (2hP+R_s)} + e^{2\pi i \frac{a}{H} (2hP+R_s)}) \sum_{n=0}^{\infty} \psi^{(2hP+R_s)}_{2HP}(n) q^{\frac{n^2}{4HP}} 	\\ 
&\hspace{0mm}	- \sum_{h=0}^{\frac{H-1}{2}-1} (e^{-2\pi i \frac{a}{H} (2(h+1)P-R_s)} + e^{2\pi i \frac{a}{H} (2(h+1)P-R_s)}) \sum_{n=0}^{\infty} \psi^{(2(h+1)P-R_s)}_{2HP}(n) q^{\frac{n^2}{4HP}} \bigg)	\label{der1}
\end{split}
\end{align}
Since the sum over $n$ is nonzero only when $n= 2HPl \pm c$ due to the presence of the periodic function $\psi^{(c)}_{2HP}(n)$, we can have
\begin{align}
\begin{split}
&\sum_{s=0}^{3} \bigg( \sum_{h=0}^{\frac{H-1}{2}}  \sum_{n=0}^{\infty} (e^{-2\pi i \frac{t}{H}n} + e^{2\pi i \frac{t}{H}n}) \psi^{(2hP+R_s)}_{2HP}(n) q^{\frac{n^2}{4HP}} - \sum_{h=0}^{\frac{H-1}{2}-1} \sum_{n=0}^{\infty} (e^{-2\pi i \frac{t}{H}n} + e^{2\pi i \frac{t}{H}n})\psi^{(2(h+1)P-R_s)}_{2HP}(n) q^{\frac{n^2}{4HP}} \bigg)	\\
&=\sum_{s=0}^{3} \sum_{n=0}^{\infty} (e^{-2\pi i \frac{t}{H}n} + e^{2\pi i \frac{t}{H}n}) \sum_{h=0}^{\frac{H-1}{2}}  \Big( \sum_{h=0}^{\frac{H-1}{2}}\psi^{(2hP+R_s)}_{2HP}(n) -\sum_{h=0}^{\frac{H-1}{2}-1}\psi^{(2(h+1)P-R_s)}_{2HP}(n) \Big) q^{\frac{n^2}{4HP}}	\label{der2}
\end{split}
\end{align}
Since
\begin{equation}
\frac{\prod_{j=1}^3 e^{y/P_j}-e^{-y/P_j}}{e^{y}-e^{-y}} = \sum_{n=0}^{\infty} \chi_{2P}(n) e^{-\frac{n}{P}y}
\end{equation}
and 
\begin{equation}
\chi_{2P}(n) = \sum_{s=0}^3 \psi_{2P}^{(R_s)}(n)	\,	,	\qquad	\psi_{2P}^{(l)}(n) = \sum_{h=0}^{\frac{H-1}{2}}\psi^{(l)}_{2HP}(n) -\sum_{h=0}^{\frac{H-1}{2}-1}\psi^{(l)}_{2HP}(n)	\,	,
\end{equation}
the Borel resummation that gives \eqref{der2} is given by
\begin{equation}
\int_{\gamma} dy e^{-\frac{K}{2\pi i} \frac{H}{P}y^2} \sum_{n=0}^{\infty} \chi_{2P}(n) e^{-\frac{n}{P}y} (e^{-2\pi i \frac{t}{H}n} + e^{2\pi i \frac{t}{H}n})	\,	.
\end{equation}
This can also be written as
\begin{align}
\begin{split}
Z_t &= \int_{\gamma} dy e^{-\frac{K}{2\pi i} \frac{H}{P}y^2} \sum_{n=0}^{\infty} \chi_{2P}(n) \Big( e^{-\frac{n}{P}(y-2\pi i \frac{P}{H}t)} + e^{-\frac{n}{P}(y+2\pi i \frac{P}{H}t)} \Big)	\\
&= \int_{\gamma} dy \Big( e^{-\frac{K}{2\pi i} \frac{H}{P}(y+2\pi i \frac{P}{H}t)^2} +e^{-\frac{K}{2\pi i} \frac{H}{P}(y+2\pi i \frac{P}{H}(H-t))^2} \Big) 
\frac{\prod_{j=1}^3 e^{y/P_j}-e^{-y/P_j}}{e^{y}-e^{-y}}	\label{bst}
\end{split}
\end{align}
since $\gamma$ is parallel to the imaginary axis of $y$-plane.
We can do similarly for the case $t=0$ and we have
\begin{align}
Z_0 = \int_{\gamma} dy e^{-\frac{K}{2\pi i} \frac{H}{P}y^2} \frac{\prod_{j=1}^3 e^{y/P_j}-e^{-y/P_j}}{e^{y}-e^{-y}}	\,	.	\label{bs0}
\end{align}
We note that \eqref{bst} and \eqref{bs0} are also obtained from \eqref{int-genh} with $\text{Im }K<0$ and with the contour $\gamma$.
Thus, from \eqref{csab}, the analytically continued CS partition function would be expressed as the Borel resummation\footnote{See also a discussion in Appendix \ref{app:h2g}.}
\begin{align}
\begin{split}
Z_{M_3}(K) &= e^{2\pi i K S_{\mathbb{0}}} \int_{\gamma} dy e^{-\frac{K}{2\pi i} \frac{H}{P}y^2} \frac{\prod_{j=1}^3 e^{y/P_j}-e^{-y/P_j}}{e^{y}-e^{-y}}	\\
&+\sum_{t=1}^{\frac{H-1}{2}} e^{2\pi i K S_{\mathbb{t}}} \int_{\gamma} dy \Big( e^{-\frac{K}{2\pi i} \frac{H}{P}(y+2\pi i \frac{P}{H}t)^2} +e^{-\frac{K}{2\pi i} \frac{H}{P}(y+2\pi i \frac{P}{H}(H-t))^2} \Big) \frac{\prod_{j=1}^3 e^{y/P_j}-e^{-y/P_j}}{e^{y}-e^{-y}}	
\end{split}	\label{ibrs}
\end{align}
when $H$ is odd where $CS(t) = \frac{P}{H} t^2$ mod 1 and $S_\mathbb{t} = CS(t) + \mathbb{Z}$.	\\

\begin{figure}
\centering
\includegraphics[width=40mm]{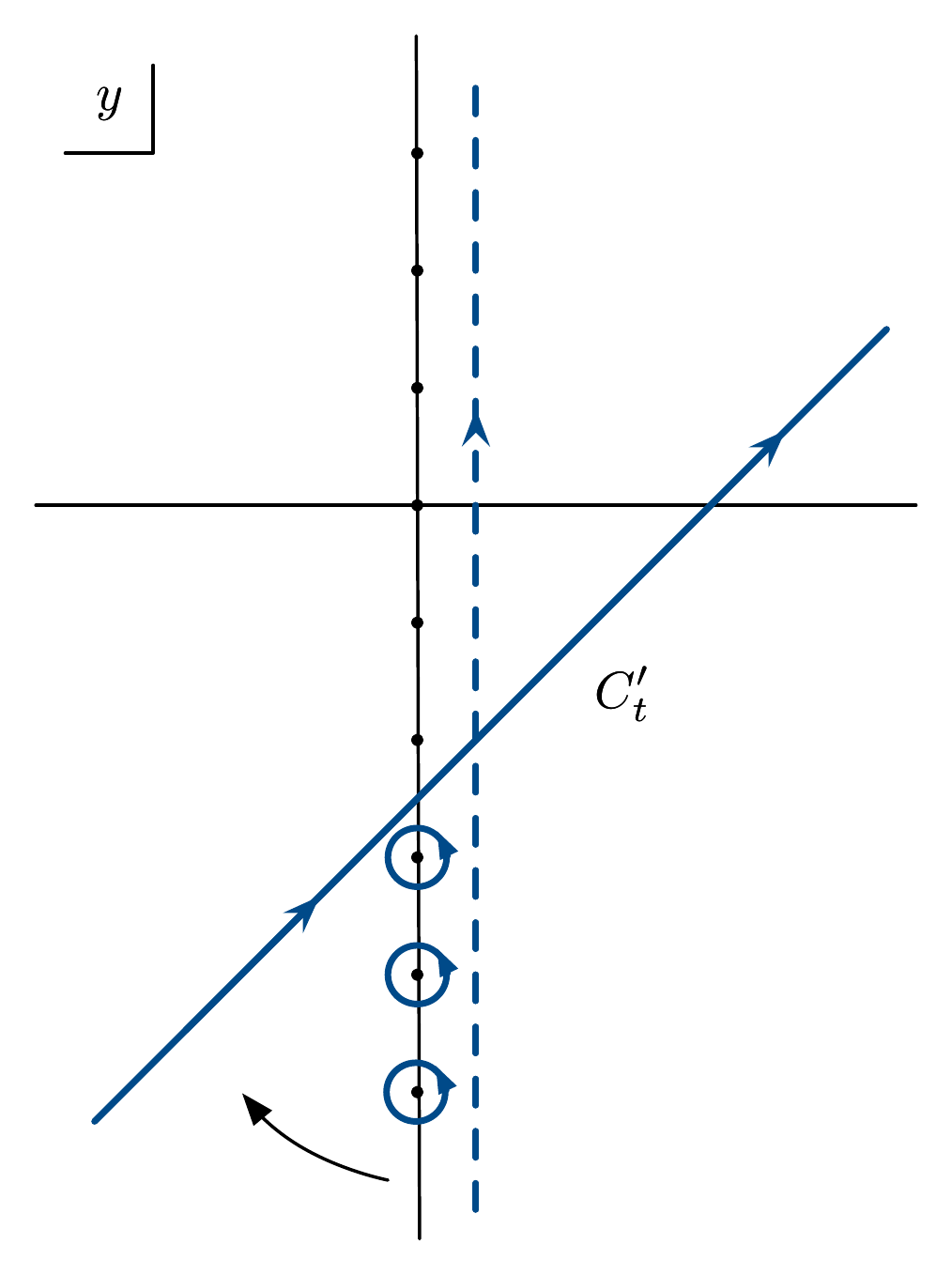} 
\caption{The case of $H \geq 2$} 
\label{Bresum4}
\end{figure}

When taking $K$ to be an integer, we change the contour from $\gamma$ to $C'_t$ that passes $y=-2 \pi i \frac{P}{H} t$ as in Figure \ref{Bresum4}.
More specifically, the part containing $e^{-\frac{K}{2\pi i} \frac{H}{P} (y+2 \pi i \frac{P}{H} t)^2}$ in \eqref{ibrs} becomes
\begin{align}
\begin{split}
&e^{2 \pi i K \frac{P}{H}t^2} \int_{C'_t} dy \, e^{-\frac{K}{2\pi i} \frac{H}{P} \big( y + 2 \pi i \frac{P}{H} t \big)^2} \frac{\prod_{j=1}^3 (e^{\frac{1}{P_j}y} - e^{-\frac{1}{P_j}y})}{e^{y} - e^{-y}}	\\
&+ 2 \pi i \sum_{m=-\infty}^{-\lfloor \frac{2Pt}{H} \rfloor -1} \text{Res } \bigg[ e^{2 \pi i K \frac{P}{H}t^2} e^{-\frac{K}{2\pi i} \frac{H}{P} \big( y + 2 \pi i \frac{P}{H} t \big)^2} \frac{\prod_{j=1}^3 (e^{\frac{1}{P_j}y} - e^{-\frac{1}{P_j}y})}{e^{y} - e^{-y}} , y = m \pi i \bigg]	
\end{split}	\label{stokeh-1}
\end{align}
After regularization, the residue part of \eqref{stokeh-1} becomes
\begin{align}
\sum_{m=1}^{2P} 4 \pi (-1)^m e^{-\frac{\pi i}{2} K \frac{H}{P} m^2} \frac{m}{P} \prod_{j=1}^3 \sin \frac{\pi m}{P_j}  
+ 8 \pi \sum_{m=1}^{\lfloor \frac{2Pt}{H} \rfloor} (-1)^m e^{-\frac{\pi i}{2}K \frac{H}{P} m^2} \prod_{j=1}^3 \sin \frac{\pi m}{P_j}	\,	.	\label{resh-1}
\end{align}
Meanwhile, upon the change of the contour from $\gamma$ to $C'_{H-t}$, the part containing $e^{-\frac{K}{2\pi i} \frac{H}{P} (y+2 \pi i \frac{P}{H} (H-t))^2}$ in \eqref{ibrs} becomes
\begin{align}
\begin{split}
&e^{2 \pi i K \frac{P}{H}t^2} \int_{C'_{H-t}} dy \, e^{-\frac{K}{2\pi i} \frac{H}{P} \big( y + 2 \pi i \frac{P}{H} (H-t) \big)^2} \frac{\prod_{j=1}^3 (e^{\frac{1}{P_j}y} - e^{-\frac{1}{P_j}y})}{e^{y} - e^{-y}}	\\
&+ 2 \pi i \sum_{m=-\infty}^{-\lfloor \frac{2P(H-t)}{H} \rfloor-1} \text{Res } \bigg[ e^{2 \pi i K \frac{P}{H}t^2} e^{-\frac{K}{2\pi i} \frac{H}{P} \big( y + 2 \pi i \frac{P}{H} (H-t) \big)^2} \frac{\prod_{j=1}^3 (e^{\frac{1}{P_j}y} - e^{-\frac{1}{P_j}y})}{e^{y} - e^{-y}} , y = m \pi i \bigg]	\,	.
\end{split}	\label{stokeh-2}
\end{align}
The residue part of \eqref{stokeh-2} is given by
\begin{align}
\sum_{m=1}^{2P} 4 \pi (-1)^m e^{-\frac{\pi i}{2} K \frac{H}{P} m^2} \frac{m}{P} \prod_{j=1}^3 \sin \frac{\pi m}{P_j}  
+ 8 \pi \sum_{m=1}^{\lfloor \frac{2P(H-t)}{H} \rfloor} (-1)^m e^{-\frac{\pi i}{2}K \frac{H}{P} m^2} \prod_{j=1}^3 \sin \frac{\pi m}{P_j}	\,	\label{resh-2}
\end{align}
after regularization.

Therefore, we see that the sum of the integral part of \eqref{stokeh-1} and \eqref{stokeh-2} is the same as the integral part of \eqref{wrt-su2-orig2} and the sum of the residues \eqref{resh-1} and \eqref{resh-2} agrees with the total residue \eqref{wrt-res-h}.
Hence, this indicates that the contributions from the non-abelian flat connections can be recovered from the contributions from the abelian flat connections via the Stokes phenomena also when $H \geq 2$.	\\

In addition, we can read off the transseries parameters for an integer $K$ from the residue parts of \eqref{stokeh-1} and \eqref{stokeh-2}.
For the examples considered in this paper, which take a form of \eqref{resurg-H1}, upon $\theta=0$ the partition function \eqref{csab} can be expressed as 
\begin{align}
Z_{M_3}(K) = \sum_{a} \Big( I_{\mathbb{a}}(|K|,0) + \frac{1}{2} \sum_{\bbbeta} m_{\mathbb{a}}^{\bbbeta} I_{\bbbeta}(|K|,0) \Big)	\,	,	\quad	m_{\mathbb{a}}^{\bbbeta} \in \mathbb{Z}	\label{spt0}
\end{align}
where $m_{\mathbb{a}}^{\bbbeta}$'s are the transseries parameters.
We provide an example for transseries parameters for an integer $K$.
We discuss the case of a general $K$ in Appendix \ref{app:h2g}.


\subsubsection*{Example}

We consider an example with $H=5$ and $(P_1,P_2,P_3) = (2,3,7)$.
In this case, the WRT invariant is given by
\begin{align}
Z_{M_{3}}(K)
=- \frac{1}{4 \sqrt{42}} e^{\frac{3}{4}\pi i} \frac{1}{2i} q^{-\frac{185}{168}} \bigg( \frac{84i}{5K} \bigg)^{\frac{1}{2}} (Z_0 + e^{\frac{4\pi i}{5}K} Z_1 + e^{\frac{6\pi i}{5}K} Z_2) \bigg|_{q \searrow e^{\frac{2\pi i}{K}}}
\end{align}
where $Z_a$'s, $a=0,1,2$, are
\begin{align}
Z_a = \pi \bigg( \frac{84i}{5K} \bigg)^{-\frac{1}{2}} \int_{\gamma} dy \, \Big( e^{-\frac{K}{2\pi i} \frac{H}{P} \big( y + 2 \pi i \frac{P}{H} a \big)^2} + e^{-\frac{K}{2\pi i} \frac{H}{P} \big( y - 2 \pi i \frac{P}{H} a \big)^2} \Big) \frac{\prod_{j=1}^F (e^{\frac{1}{P_j}y} - e^{-\frac{1}{P_j}y})}{e^{y} - e^{-y}}	\,	.	\label{za-237-h5}
\end{align}
$a=0$ corresponds to the trivial flat connection, and $a=1,2$ correspond to the abelian flat connections.
$Z_a$ is expressed in terms of homological block $\widehat{Z}_b$ as
\begin{align}
Z_a = \sum_{b=0}^{2} S_{ab} \widehat{Z}_b
\end{align}
where 
\begin{align}
\begin{split}
\widehat{Z}_0 = \widetilde{\Psi }_{210}^{(55)}+\widetilde{\Psi }_{210}^{(85)}+\widetilde{\Psi }_{210}^{(125)}+\widetilde{\Psi }_{210}^{(155)}	\,	,
\end{split}	\\
\begin{split}
\widehat{Z}_1 = \widetilde{\Psi }_{210}^{(1)}-\widetilde{\Psi }_{210}^{(29)}+\widetilde{\Psi }_{210}^{(41)}+\widetilde{\Psi }_{210}^{(71)}+\widetilde{\Psi }_{210}^{(139)}+\widetilde{\Psi}_{210}^{(169)}+\widetilde{\Psi }_{210}^{(209)}+\widetilde{\Psi }_{210}^{(239)}	\,	,
\end{split}	\\
\begin{split}
\widehat{Z}_2 = -\widetilde{\Psi }_{210}^{(13)}-\widetilde{\Psi }_{210}^{(43)}-\widetilde{\Psi }_{210}^{(83)}-\widetilde{\Psi }_{210}^{(97)}-\widetilde{\Psi }_{210}^{(113)}-\widetilde{\Psi}_{210}^{(127)}-\widetilde{\Psi }_{210}^{(167)}+\widetilde{\Psi }_{210}^{(223)}	\,	,
\end{split}
\end{align}
and
\begin{align}
S_{ab} = \frac{1}{\sqrt{5}}
\begin{pmatrix}
1	&	1					&	1	\\
2	&	\frac{1}{2}(\sqrt{5}-1)		&	\frac{1}{2}(-\sqrt{5}-1)	\\
2	&	\frac{1}{2}(-\sqrt{5}-1)	&	\frac{1}{2}(\sqrt{5}-1)	
\end{pmatrix}	\,	.
\end{align}

When $K \in \mathbb{Z}_+$, we change the integration contour as in Figure \ref{Bresum4}.
There are three sets of poles at $y=-n \pi i$ in \eqref{za-237-h5} with
\begin{align}
n =
\begin{cases}
1, 13, 29, 41, 43, 55, 71, 83	\\
5, 19, 23, 37, 47, 61, 65, 79	\\
11, 17, 25, 31, 53, 59, 67, 73
\end{cases}
\hspace{3mm} \text{mod } 84	\,	.
\end{align}
In this case, it is expected that there are three $SL(2,\mathbb{C})$ non-abelian flat connections, which we denote as $\alpha_1$, $\alpha_2$, and $\alpha_3$, and
their Chern-Simons invariants are
\begin{align}
CS(\alpha_1) = -\frac{5}{168}	\,	,	\qquad
CS(\alpha_2) = -\frac{125}{168}	\,	,	\qquad
CS(\alpha_3) = -\frac{101}{168}	\,	,	\qquad \text{mod } 1	\,	,
\end{align}
respectively.
This case is rather a special case due to the condition that $P_j$'s are pairwise coprime.\footnote{More specifically, in general the building block for the homological block of Seifert manifolds with $F=3$ when $H \neq 1$ is $\widetilde{\Psi}_{HP}^{(a)}(q)$ and the integrand of the corresponding Borel resummation contains a rational function of sine hyperbolic functions that depend on $H$.
However, at least when $P_j$'s are pairwise coprime, the building blocks $\widetilde{\Psi}_{HP}^{(a)}(q)$ are summed up in such a way that $Z_t$ contains a rational function of sine hyperbolic functions that doesn't depend on $H$.

As a side remark, though we consider an example from an infinite family of Seifert manifolds where $P_j$'s are coprime, a similar analysis can be done for the examples that the expression in terms of $\widetilde{\Psi}_{HP}^{(a)}(q)$ are known, such as examples discussed in \cite{Cheng-Chun-Ferrari-Gukov-Harrison}, by using \eqref{resurg-CS}.
}

From \eqref{resh-1} and \eqref{resh-2}, the contributions from the non-abelian flat connections attached to the abelian flat connection $a$ are
\begin{empheq}{alignat*=3}
a_0: a=0	&\quad\leadsto	&&\quad	-8 \pi  \sqrt{3} e^{\frac{43 \pi i K}{84}} \cos \left(\frac{3 \pi }{14}\right)-8 \pi  \sqrt{3} e^{\frac{67 \pi i K}{84}} \cos \left(\frac{\pi }{14}\right)	\\
a_1: a=1	&\quad\leadsto	&&\quad	-8 \pi  \sqrt{3} e^{\frac{43 \pi i K}{84}} \cos \left(\frac{3 \pi }{14}\right)-8 \pi  \sqrt{3} e^{\frac{67 \pi i K}{84}} \cos \left(\frac{\pi }{14}\right)	\\
a_2: a=2	&\quad\leadsto 	&&\quad	8 \pi  \sqrt{3} e^{\frac{163 \pi i K}{84}} \sin \left(\frac{\pi }{7}\right)+8 \pi  \sqrt{3} e^{\frac{43 \pi i K}{84}} \cos \left(\frac{3 \pi }{14}\right)+16 \pi  \sqrt{3} e^{\frac{67 \pi i K}{84}} \cos \left(\frac{\pi }{14}\right)	\,	,
\end{empheq}
so the total sum of the residues or the total contribution from the non-abelian flat connections is
\begin{align}
\sum_{\alpha = \alpha_1,\alpha_2,\alpha_3} n_\alpha e^{2\pi i K CS(\alpha)} Z_{\alpha}
= 8 \sqrt{3} \pi  e^{\frac{163 \pi i K}{84}} \sin \left(\frac{\pi }{7}\right) -8 \sqrt{3} \pi  e^{\frac{43 \pi i K}{84}} \cos \left(\frac{3 \pi }{14}\right)	\,	.	\label{tot-nonab-int-237}
\end{align}
where $Z_{\alpha_1} = 8 \sqrt{3} \pi \sin \left(\frac{\pi }{7}\right)$, $Z_{\alpha_2}= -8 \sqrt{3} \pi \cos \left(\frac{3 \pi }{14}\right)$, and $Z_{\alpha_3}=-8 \sqrt{3} \pi  \cos \left(\frac{\pi }{14}\right)$.
Thus, $\alpha_1$ and $\alpha_2$ are real non-abelian flat connections, $n_{\alpha_1}=n_{\alpha_2}=1$, while $\alpha_3$ is a complex non-abelian flat connection, $n_{\alpha_3}=0$.
We also checked this via the modularity discussed in \cite{Cheng-Chun-Ferrari-Gukov-Harrison}.	\\

From \eqref{stokeh-1} and \eqref{stokeh-2}, we can obtain transseries parameters associated to $y=-n\pi i$ for each abelian flat connections $a$ when $K \in \mathbb{Z}_+$,
\begin{empheq}[left={m_{(a_0,0)}^{\bbbeta} = \empheqlbrace}]{alignat*=2}
&\ \text{for } \bbbeta=(\alpha_1,-\frac{5}{168}n^2)	&&\left\{ 
\begin{array}{l l l l l }
+1	&\quad n = 1, 41, 55, 71	&\ \text{mod } 84	\\
-1	&\quad n = 13, 29, 43, 83	&\ \text{mod } 84	
\end{array}	\right.			\\
&\ \text{for } \bbbeta=(\alpha_2,-\frac{5}{168}n^2)	&&\left\{ 
\begin{array}{l l l l l }
+1	&\quad n = 5, 19, 23, 37	&\ \text{mod } 84	\\
-1	&\quad n = 47, 61, 65, 79	&\ \text{mod } 84	
\end{array}	\right.			\\
&\ \text{for } \bbbeta=(\alpha_3,-\frac{5}{168}n^2)	&&\left\{ 
\begin{array}{l l l l l }
+1	&\quad n = 11, 17, 25, 31	&\ \text{mod } 84	\\
-1	&\quad n = 53, 59, 67, 73	&\ \text{mod } 84	
\end{array}	\right.		
\end{empheq}

\begin{empheq}[left={m_{(a_1,\frac{2}{5})}^{\bbbeta} = \empheqlbrace}]{alignat*=2}
&\ \text{for } \bbbeta=(\alpha_1,-\frac{5}{168}n^2)	&&\left\{ 
\begin{array}{l l l l l }
-1	&\quad n = 29, 43		&\				\\
+1	&\quad n = 41, 55		&\ 				\\
+2	&\quad n = 71			&\ \text{mod } 84	\\
-2	&\quad n = 83			&\ \text{mod } 84	\\
+2	&\quad n = 1, 41, 55		&\ \text{mod } 84	\text{ with } n > 84		\\
-2	&\quad n = 13,29,43		&\ \text{mod } 84	\text{ with } n > 84	
\end{array}	\right.			\\
&\ \text{for } \bbbeta=(\alpha_2,-\frac{5}{168}n^2)	&&\left\{ 
\begin{array}{l l l l l }
+1	&\quad n = 19, 23, 37	&\				\\
-1	&\quad n = 47, 61, 65	&\ 				\\
-2	&\quad n = 79			&\ \text{mod } 84	\\
+2	&\quad n = 5, 19, 23, 37	&\ \text{mod } 84	\text{ with } n > 84	\\
-2	&\quad n = 47, 61, 65	&\ \text{mod } 84	\text{ with } n > 84
\end{array}	\right.			\\
&\ \text{for } \bbbeta=(\alpha_3,-\frac{5}{168}n^2)	&&\left\{ 
\begin{array}{l l l l l }
+1	&\quad n = 17, 25, 31	&\				\\
-1	&\quad n = 53, 59, 67	&\ 				\\
-2	&\quad n = 73			&\ \text{mod } 84	\\
+2	&\quad n = 11, 17, 25, 31	&\ \text{mod } 84	\text{ with } n > 84	\\
-2	&\quad n = 53, 59, 67	&\ \text{mod } 84	\text{ with } n > 84
\end{array}	\right.		
\end{empheq}

\begin{empheq}[left={m_{(a_2,\frac{3}{5})}^{\bbbeta} = \empheqlbrace}]{alignat*=2}
&\ \text{for } \bbbeta=(\alpha_1,-\frac{5}{168}n^2)	&&\left\{ 
\begin{array}{l l l l l }
+1	&\quad n = 41			&\				\\
-1	&\quad n = 43			&\				\\
+2	&\quad n = 55, 71		&\ \text{mod } 84	\\
-2	&\quad n = 83			&\ \text{mod } 84	\\
+2	&\quad n = 1, 41		&\ \text{mod } 84	\text{ with } n > 84		\\
-2	&\quad n = 13, 29, 43	&\ \text{mod } 84	\text{ with } n > 84	
\end{array}	\right.			\\
&\ \text{for } \bbbeta=(\alpha_2,-\frac{5}{168}n^2)	&&\left\{ 
\begin{array}{l l l l l }
+1	&\quad n = 37			&\				\\
-1	&\quad n = 47			&\ 				\\
-2	&\quad n = 61, 65, 79	&\ \text{mod } 84	\\
+2	&\quad n = 5, 19, 23, 37	&\ \text{mod } 84	\text{ with } n > 84	\\
-2	&\quad n = 47			&\ \text{mod } 84	\text{ with } n > 84
\end{array}	\right.			\\
&\ \text{for } \bbbeta=(\alpha_3,-\frac{5}{168}n^2)	&&\left\{ 
\begin{array}{l l l l l }
-2	&\quad n = 53, 59, 67, 73		&\ \text{mod } 84	\\
+2	&\quad n = 11, 17, 25, 31		&\ \text{mod } 84	\text{ with } n > 84
\end{array}	\right.		
\end{empheq}
With $n_\alpha = \sum_{a} \sum_{\beta} \frac{1}{2} m_{\mathbb{a}}^{\, \bbbeta}$, we see that $n_{\alpha_1}=n_{\alpha_2}=1$ and $n_{\alpha_3}=0$ from the results above.


\section{Resurgent analysis for Seifert manifolds with rational level $K$}
\label{sec:rationalk}

We are also interested in the resurgent analysis when the level $K$ is taken to be a rational number.

The full $G=SU(2)$ WRT invariant for the Seifert manifolds $X(P_1/Q_1,\ldots, P_F/Q_F)$ when the level $K=\frac{r}{s}$ is a rational number with coprime $r$ and $s$ \cite{Lawrence-Rozansky}\footnote{More precisely, for the examples discussed in this paper, the condition is that $s$ is coprime to $4KP$, which was used in \cite{Lawrence-Rozansky} to derive \eqref{wrt-su2-rat} via the Galois action on $e^{2\pi i \frac{1}{4KP}}$.} can be expressed as
\begin{align}
\begin{split}
\sum_{t=0}^{Hs-1} \int_{C_0'} f(y) e^{K g_t(y)} dy 
- 2 \pi i \sum_{m=1}^{2Ps-1} \text{Res} \bigg( \frac{f(y) e^{K g_0(y)}}{1-e^{-2Ky}}, y=\pi i m \bigg)
\end{split}	\label{wrt-su2-rat}
\end{align}
up to an overall factor.
Here, $f(y) = \frac{\prod_{j=1}^F (e^{\frac{1}{P_j}y} - e^{-\frac{1}{P_j}y})}{(e^{y} - e^{-y})^{F-2}}$ and $g_t(y) = -\frac{1}{2\pi i}\frac{H}{P}y^2 - 2ty$ are the same as in the previous section.
Upon the change of the contour from $C_0'$ to $C_t'$ that passes a stationery phase point $y=-2 \pi i \frac{P}{H}t$ for each $t$, \eqref{wrt-su2-rat} can be expressed as
\begin{align}
\begin{split}
& \sum_{t=0}^{Hs-1} \int_{C_t'} f(y) e^{K g_t(y)} dy 	\\
&- 2 \pi i \sum_{m=1}^{2Ps-1} \text{Res} \bigg( \frac{f(y) e^{K g_0(y)}}{1-e^{-2Ky}}, y=\pi i m \bigg) 
- 2 \pi i \sum_{t=1}^{Hs-1} \sum_{m=1}^{\lfloor \frac{2Pt}{H} \rfloor} \text{Res} \Big( f(y) e^{K g_t(y)}, y = -\pi i m \Big)
\end{split}	\label{wrt-su2-rat2}
\end{align}

We first consider the residue part of \eqref{wrt-su2-rat} or the first type of the residue in the second line of \eqref{wrt-su2-rat2}, when $F=3$.
The case of other singular fibers can also be done similarly.
When $m$ is not a multiple of $s$, there is no pole from $(1-e^{-2Ky})^{-1}$ if $K= \frac{r}{s}$, but the integrand can have poles of order 1 from $(e^y - e^{-y})^{-1}$ of $f(y)$.
In this case, the sum of the residues from such poles is zero.
Meanwhile, if $m$ is a multiple of $s$, the integrand has poles of order 2 from $(e^y - e^{-y})^{-1} (1-e^{-2Ky})^{-1}$ for $K=\frac{r}{s}$.
The contributions from such poles are given by
\begin{align}
-2 \pi i \sum_{n=0}^{2P-1} 2i (-1)^{ns} H e^{-\frac{\pi i}{2} \frac{H}{P} rsn^2} \Big( \frac{sn}{P} - \frac{1}{H} \Big) \prod_{j=1}^{3} \sin \Big( \frac{\pi n s}{P_j} \Big)	\,	,	\label{dpole}
\end{align}
which can be simplified to
\begin{align}
4 \pi \sum_{n=0}^{2P-1} (-1)^{ns} H e^{-\frac{\pi i}{2} \frac{H}{P} rsn^2} \frac{sn}{P} \prod_{j=1}^{3} \sin \Big( \frac{\pi n s}{P_j} \Big)	\,	.	\label{ratres1}
\end{align}
The second type of residues in \eqref{wrt-su2-rat2} is given by
\begin{align}
8 \pi \sum_{t=1}^{Hs-1} \sum_{m=1}^{\lfloor \frac{2Pt}{H} \rfloor} e^{-\frac{1}{2 \pi i} \frac{H}{P} \frac{r}{s} (-m \pi i + 2 \pi i \frac{P}{H} t)^2 } e^{ 2 \pi i \frac{r}{s} \frac{P}{H} t^2} (-1)^m \prod_{j=1}^{3} \sin \frac{\pi m}{P_j}		\,	.
\label{ratres2}
\end{align}
Therefore, the total residue is given by the sum of \eqref{ratres1} and \eqref{ratres2},
\begin{align}
4 \pi \sum_{n=0}^{2P-1} (-1)^{ns} H e^{-\frac{\pi i}{2} \frac{H}{P} rsn^2} \frac{sn}{P} \prod_{j=1}^{3} \sin \Big( \frac{\pi n s}{P_j} \Big)
+\sum_{t=1}^{Hs-1} \sum_{m=1}^{\lfloor \frac{2Pt}{H} \rfloor} 8 \pi (-1)^m e^{-\frac{\pi i}{2} \frac{H}{P} K m^2 + 2 \pi i K m t} \prod_{j=1}^{3} \sin \frac{\pi m}{P_j}	\,	.	\label{rtotres}
\end{align}


\subsection*{Resurgent analysis}

As in the case of an integer $K$, by analytically continuing the level $K$ from a rational $K$ and choosing a contour $\gamma$ in the Gaussian integral part of \eqref{wrt-su2-rat}, we can see that the WRT invariant can be expressed as
\begin{align}
\begin{split}
&\hspace{-15mm}\text{WRT}_{M_3}(K=r/s) \simeq \sum_{v=0}^{s-1} \bigg( e^{2\pi i \frac{r}{s} HP v^2}  \sum_{m=0}^{3} \Big(  \sum_{h=0}^{ \frac{H-1}{2} }  \widetilde{\Psi}^{(2hP+R_m)}_{HP}(q) - \sum_{h=0}^{ \frac{H-1}{2}-1 }  \widetilde{\Psi}^{(2(h+1)P-R_m)}_{HP}(q) \Big) \bigg)	\\
&\hspace{5mm} + \sum_{u=1}^{\frac{H-1}{2}} \sum_{v=0}^{s-1} 
\bigg[ 
e^{2\pi i \frac{r}{s} \frac{P}{H} (Hv+u)^2} 
\sum_{m=0}^{3} \bigg(  
\sum_{h=0}^{ \frac{H-1}{2} } \big(  e^{-2\pi i \frac{u}{H} (2hP+R_m)} + e^{2\pi i \frac{u}{H} (2hP+R_m)} \big) \widetilde{\Psi}^{(2hP+R_m)}_{HP}(q)	\\
&\hspace{35mm}- \sum_{h=0}^{ \frac{H-1}{2}-1 } \big( e^{-2\pi i \frac{u}{H} (2(h+1)P-R_m)} + e^{2\pi i \frac{u}{H} (2(h+1)P-R_m)} \big) \widetilde{\Psi}^{(2(h+1)P-R_m)}_{HP}(q) \Big) 	
\bigg]	\bigg|_{q \searrow e^{2\pi i \frac{s}{r}}}	\,	,	\label{wrt-rs-o}
\end{split}	
\end{align}
up to an overall factor, when $H$ is odd and $F=3$ where $R_j$, $j=0,1,2,3$, are given in \eqref{rj} \cite{Chung-rationalk}.\footnote{See also \cite{Kucharski:2019fgh} for the expression of the WRT invariant for a rational $K$ in terms of homological blocks for plumbed 3-manifolds.}	
We note that when deriving \eqref{wrt-rs-o} we took $t=Hv+u$ and expressed the sum over $t=0,1, \ldots, Hs-1$ as the sum over $u=0,1, \ldots, H-1$ and $v=0,1, \ldots, s-1$.

The structure of \eqref{wrt-rs-o} takes a form of 
\begin{align}
\text{WRT}_{M_3}(K)\simeq \sum_{v=0}^{s-1} \sum_{u=0}^{\frac{H-1}{2}} e^{2\pi i \frac{r}{s} \frac{P}{H} (Hv+u)^2 } Z_u(q)	\,	\Big|_{q \searrow e^{2\pi i \frac{s}{r}}}	\,	,	\qquad	Z_u(q) = \sum_{b=0}^{\frac{H-1}{2}} S_{ub} \widehat{Z}_b(q)	\label{wrt-rat}
\end{align}
where $Z_u$'s are given by \eqref{z0} and \eqref{zt}.
We see that the differences from the case of the integer $K$ are the factor $e^{2\pi i \frac{r}{s} \frac{P}{H} (vH+u)^2}$ and the summation $\sum_{v=0}^{s-1}$.
The structure \eqref{wrt-rat} tells that $u$ labels the abelian flat connection as in the case of integer $K$.
In this section, we discuss the case of an odd $H$ and the case of even $H$ can be done similarly.	\\

The structure \eqref{wrt-rat} at a rational $K$ is different from \eqref{csab} with \eqref{ia} or its limit when $K \in \mathbb{Z}_+$ in that \eqref{wrt-rat} is expressed as a sum over $v$ but there is no such a sum in \eqref{csab}.
As discussed in \cite{Witten-accs}, a lift for a flat connection is chosen in the analytically continued Chern-Simons partition function as in \eqref{csab}.
Also, by construction, when $K$ is taken to be an integer, \eqref{csab} with \eqref{ia} or \eqref{spt0} becomes the standard path integral of Chern-Simons theory for an integer $K$.
In our discussion, when $K \in \mathbb{Z}_+$, any value of $S_{\mathbb{a}} \in CS(a) + \mathbb{Z}$ that captures a lift gives $e^{2\pi i K CS(a)}$ for an abelian flat connection $a$, and there is no sum like \eqref{wrt-rat}.

From a viewpoint of the analytically continued Chern-Simons theory, the case of a rational $K$ can be regarded as another limit of an analytic continuation from an integer $K$.
For a general complex $K$ such that $|q|<1$ (or $|q|>1$), the construction of an analytically continued Chern-Simons theory doesn't necessarily depend on whether $K=r$ or $K= \frac{r}{s}$ for $r \in \mathbb{C}$, since they are just one of the value of $K$ with $|q|<1$.
Thus, it would not be natural to expect that there are two types of expression of the analytically continued Chern-Simons partition function obtained from the analytic continuation from $K \in \mathbb{Z}$ and from $K \in \mathbb{Q}$.
Rather, it would be appropriate that for a general value of $K \in \mathbb{C}$ an analytic continued Chern-Simons partition function would be given by \eqref{csab} with \eqref{ia} and an expression as a sum over lifts\footnote{A large gauge transformation that shifts the Chern-Simons invariant by the amount of $s$ would be a symmetry.
Accordingly, the number of lifts in the sum would be at most $s$.
Meanwhile, considering the value of $e^{2\pi i \frac{r}{s} \frac{P}{H} (Hv+u)^2 }$, $v=0, 1, \ldots, s-1$, we see that there are $\frac{s+1}{2}$ number of distinct exponents $\frac{r}{s} \frac{P}{H} (Hv+u)^2$ mod 1 given a $u$ where $s$ is odd, which may imply that there are $\frac{s+1}{2}$ number of lifts.
} as in \eqref{wrt-rat} arise at $K \in \mathbb{Q}$. 
\eqref{wrt-rs-o} or \eqref{wrt-rat} tell how the sum over lifts of abelian flat connections arise explicitly in the WRT invariant of the Seifert manifolds at $K=\frac{r}{s}$, but we don't have an explanation on why the sum of exponential factors in \eqref{wrt-rat} takes such a form precisely.	\\

As done previously, when taking $K=\frac{r}{s}$, we change the contour from $\gamma$ to $C'_{u}$ and $C'_{H-u}$, and we have
\begin{align}
\begin{split}
\sum_{v=0}^{s-1} &\Bigg( e^{2 \pi i \frac{r}{s} \frac{P}{H}v^2} \int_{C'_0} dy \, e^{-\frac{1}{2\pi i} \frac{r}{s} \frac{H}{P} y^2} \frac{\prod_{j=1}^3 (e^{\frac{1}{P_j}y} - e^{-\frac{1}{P_j}y})}{e^{y} - e^{-y}}	\\
&+\sum_{u=1}^{\frac{H-1}{2}} e^{2 \pi i \frac{r}{s} \frac{P}{H}(Hv+u)^2} \bigg( \int_{C'_{u}} dy \, e^{-\frac{1}{2\pi i} \frac{r}{s} \frac{H}{P} \big( y + 2 \pi i \frac{P}{H} u \big)^2} \frac{\prod_{j=1}^3 (e^{\frac{1}{P_j}y} - e^{-\frac{1}{P_j}y})}{e^{y} - e^{-y}}	\\
&\hspace{40mm} + \int_{C'_{H-u}} dy \, e^{-\frac{1}{2\pi i} \frac{r}{s} \frac{H}{P} \big( y + 2 \pi i \frac{P}{H} (H-u) \big)^2} \frac{\prod_{j=1}^3 (e^{\frac{1}{P_j}y} - e^{-\frac{1}{P_j}y})}{e^{y} - e^{-y}} \bigg)  
\Bigg)
\end{split}	\label{irsr-int}
\end{align}
for the integral part.
We can see that \eqref{irsr-int} agrees with the integral part of \eqref{wrt-su2-rat2}.

For the residue part, we obtain
\begin{align}
\begin{split}
&\hspace{-17mm}8 \pi \sum_{v=0}^{s-1} \Bigg( e^{2 \pi i \frac{r}{s} \frac{P}{H}(Hv)^2} \sum_{m=-\infty}^{0} e^{-\frac{\pi}{2} \frac{r}{s} \frac{H}{P} m^2} (-1)^m \prod_{j=1}^3 \sin \frac{\pi m}{P_j}	\\
&\hspace{-0mm}+\sum_{u=1}^{\frac{H-1}{2}} e^{2 \pi i \frac{r}{s} \frac{P}{H}(Hv+u)^2} \bigg( \sum_{m=-\infty}^{-\lfloor \frac{2Pu}{H} \rfloor -1} e^{-\frac{\pi}{2} \frac{r}{s} \frac{H}{P} \big( m + \frac{2Pu}{H}\big)^2} (-1)^m \prod_{j=1}^3 \sin \frac{\pi m}{P_j}	\\
&\hspace{45mm}+ \sum_{m=-\infty}^{-\lfloor \frac{2P(H-u)}{H} \rfloor-1} e^{-\frac{\pi}{2} \frac{r}{s} \frac{H}{P} \big( m + \frac{2P(H-u)}{H}\big)^2} (-1)^m \prod_{j=1}^3 \sin \frac{\pi m}{P_j} \bigg)
\Bigg)	\,	,
\end{split}	\label{irsr-res}
\end{align}
which becomes
\begin{align}
\begin{split}
&\hspace{-20mm}-8 \pi \sum_{v=0}^{s-1} \Bigg( 
\frac{1}{2} e^{2 \pi i \frac{r}{s} \frac{P}{H}(Hv)^2} \sum_{m=0}^{2Ps} e^{-\frac{\pi}{2} \frac{r}{s} \frac{H}{P} m^2} (-1)^m \Big( 1-\frac{m}{Ps} \Big) \prod_{j=1}^3 \sin \frac{\pi m}{P_j}	\\
&+\sum_{u=1}^{\frac{H-1}{2}} e^{2 \pi i \frac{r}{s} \frac{P}{H}(Hv+u)^2} \bigg( \frac{1}{2} \sum_{m=0}^{2Ps} \Big( e^{-\frac{\pi}{2} \frac{r}{s} \frac{H}{P} (m-\frac{2Pu}{H})^2} + e^{-\frac{\pi}{2} \frac{r}{s} \frac{H}{P} (m+\frac{2Pu}{H})^2} \Big) (-1)^m \Big( 1-\frac{n}{Ps} \Big) \prod_{j=1}^3 \sin \frac{\pi m}{P_j}	\\
&\hspace{50mm}-2 \sum_{m=0}^{\lfloor \frac{2Pu}{H} \rfloor} e^{-\frac{\pi}{2} \frac{r}{s} \frac{H}{P} (m-\frac{2Pu}{H})^2} (-1)^m \prod_{j=1}^3 \sin \frac{\pi m}{P_j} \bigg)	
\Bigg)
\label{irsr-res-reg}
\end{split}
\end{align}
after regularization.
We can also see that \eqref{irsr-res-reg} agrees with the total residue \eqref{rtotres} above.
More specifically, \eqref{ratres2} can be written as 
\begin{align}
\begin{split}
&8 \pi \sum_{t=0}^{Hs-1}\sum_{m=1}^{\infty} e^{-\frac{1}{2 \pi i} \frac{H}{P} \frac{r}{s} (-m \pi i + 2 \pi i \frac{P}{H} t)^2 } e^{ 2 \pi i \frac{r}{s} \frac{P}{H} t^2} (-1)^m \prod_{j=1}^{3} \sin \frac{\pi m}{P_j}	\\
&- 8 \pi \sum_{t=0}^{Hs-1} \sum_{m=\lfloor \frac{2Pt}{H} \rfloor+1}^{\infty} e^{-\frac{1}{2 \pi i} \frac{H}{P} \frac{r}{s} (-m \pi i + 2 \pi i \frac{P}{H} t)^2 } e^{ 2 \pi i \frac{r}{s} \frac{P}{H} t^2} (-1)^m \prod_{j=1}^{3} \sin \frac{\pi m}{P_j}	
\end{split} \label{ratres2-1}
\end{align}
and the first line of \eqref{ratres2-1} after regularization cancels out the contribution \eqref{ratres1}.
Thus, \eqref{rtotres} can be expressed as
\begin{align}
- 8 \pi \sum_{t=0}^{Hs-1} \sum_{m=\lfloor \frac{2Pt}{H} \rfloor+1}^{\infty} e^{-\frac{1}{2 \pi i} \frac{H}{P} \frac{r}{s} (-m \pi i + 2 \pi i \frac{P}{H} t)^2 } e^{ 2 \pi i \frac{r}{s} \frac{P}{H} t^2} (-1)^m \prod_{j=1}^{3} \sin \frac{\pi m}{P_j}	\,	.	\label{rattotres}
\end{align}
With $t=Hv+u$, \eqref{rattotres} is given by
\begin{align}
- 8 \pi \sum_{v=0}^{s-1} \sum_{u=0}^{H-1} e^{ 2 \pi i \frac{r}{s} \frac{P}{H} (Hv+u)^2} \sum_{m=\lfloor \frac{2Pu}{H} \rfloor+1}^{\infty} e^{-\frac{\pi i}{2} \frac{H}{P} \frac{r}{s} (-m + \frac{2Pu}{H} )^2 }  (-1)^m \prod_{j=1}^{3} \sin \frac{\pi m}{P_j}	\,	.	\label{rattotres-1}
\end{align}
After some calculations, \eqref{rattotres-1} can be expressed as
\begin{align}
\begin{split}
&\hspace{-20mm}-8 \pi \sum_{v=0}^{s-1} \Bigg( e^{2\pi i \frac{r}{s} \frac{P}{H} (Hv)^2} \sum_{m=0}^{\infty} e^{-\frac{\pi i}{2} \frac{r}{s} \frac{H}{P} m^2 } (-1)^m \prod_{j=1}^3 \sin \frac{\pi m}{P_j}	\\
&+\sum_{u=1}^{\frac{H-1}{2}} e^{2\pi i \frac{r}{s} \frac{P}{H} (Hv+u)^2}
\bigg(
\sum_{m=0}^{\infty} \big( e^{-\frac{\pi i}{2} \frac{r}{s} \frac{H}{P} \big(m-\frac{2Pu}{H}\big)^2 } + e^{-\frac{\pi i}{2} \frac{r}{s} \frac{H}{P} \big(m+\frac{2Pu}{H}\big)^2 } \big) (-1)^m \prod_{j=1}^3 \sin \frac{\pi m}{P_j}	\\
&\hspace{50mm}-2 \sum_{m=0}^{\lfloor \frac{2Pu}{H} \rfloor} e^{-\frac{\pi i}{2} \frac{r}{s} \frac{H}{P} \big(m-\frac{2Pu}{H}\big)^2 } (-1)^m \prod_{j=1}^3 \sin \frac{\pi m}{P_j}
\bigg)
\Bigg)	\,	,
\end{split}
\end{align}
and after regularization we see that this agrees with \eqref{irsr-res-reg} obtained from the Stokes phenomena.

The set of poles that give, for example, the same $e^{2\pi i K \frac{P}{H} u^2} e^{-\frac{\pi i}{2}K \frac{H}{P} (m-\frac{2Pu}{H})^2}$ in \eqref{irsr-res} when $K$ is an integer split to the poles that give the same $e^{2\pi i \frac{r}{s} \frac{P}{H} (Hv+u)^2} e^{-\frac{\pi i}{2} \frac{r}{s} \frac{H}{P} (m-\frac{2Pu}{H})^2}$ when $K$ is a rational number, $K=\frac{r}{s}$.
Given a $u$ and a $v$, we see that there are $\frac{s+1}{2}$ number of distinct values of $e^{2\pi i \frac{r}{s} \frac{P}{H} (Hv+u)^2} e^{-\frac{\pi i}{2} \frac{r}{s} \frac{H}{P} (m-\frac{2Pu}{H})^2}$ where $s$ is odd for the examples that we work here.	
Thus, given a $u$ and a $v$ there is also a sum over lifts of non-abelian flat connections.	\\

When $H=1$, \eqref{wrt-rs-o} is simplified to
\begin{align}
Z_{M_3}(K=r/s) \simeq g(Pr;s) \sum_{j=1}^{3} \widetilde{\Psi}_{2P}^{(R_j)}(q)  \bigg|_{q \searrow e^{2\pi i \frac{s}{r}}}	\,		\label{wrt-rs-h1}
\end{align}
up to an overall factor where $g(m;s)$ denotes the quadratic Gauss sum
\begin{align}
g(m;s) = \sum_{n=0}^{s-1} e^{2\pi i m n^2 / s}	\,	.
\end{align}
and also \eqref{irsr-res-reg} can be simplified to
\begin{align}
-4 \pi g(Pr;s) \sum_{m=1}^{2Ps-1} e^{-\frac{\pi i}{2} \frac{1}{P} \frac{r}{s} n^2} (-1)^m \bigg( 1-\frac{m}{Ps} \bigg) \prod_{j=1}^{3} \sin \frac{\pi m}{P_j}	\,	.	\label{rattotresH1-reg}
\end{align}
\\

As a consistency check, we see that \eqref{rattotresH1-reg} also agrees with the non-perturbative part obtained from the modularity.
More specifically, from the modularity, the non-perturbative part of $\widetilde{\Psi}_{P}^{(a)}(q=e^{2\pi i \frac{s}{r}})$ is given by
\begin{align}
-\sqrt{\frac{r/s}{i}} \sum_{b=1}^{P-1} M_{ab} \widetilde{\Psi}_{P}^{(b)}(e^{-2\pi i \frac{r}{s}}) 	\,	.	\label{rat-nonpert}
\end{align}
From the regularization, $\widetilde{\Psi}_{P}^{(b)}(e^{-2\pi i \frac{r}{s}})$ is given by $\widetilde{\Psi}_{P}^{(b)}(e^{-2\pi i \frac{r}{s}})=\frac{1}{2} \sum_{n=1}^{2Ps-1} \psi_{2P}^{(b)}(n) e^{-2 \pi i \frac{1}{4P} \frac{r}{s} n^2} \big( 1 - \frac{n}{Ps} \big)$, so \eqref{rat-nonpert} is
\begin{align}
- \frac{1}{2}\sqrt{\frac{r/s}{i}} \sum_{b=1}^{P-1} \sqrt{\frac{2}{P}} \sin \frac{\pi a b}{P} \sum_{n=1}^{2Ps-1} \psi_{2P}^{(b)}(n) e^{-2 \pi i \frac{1}{4P} \frac{r}{s} n^2} \Big( 1 - \frac{n}{Ps} \Big)	\,	.
\end{align}
When $F=3$ and $H=1$, the WRT invariant is given by $Z_{M_3}(K=\frac{r}{s}) \simeq g(Pr;s) \sum_{j=0}^3 \widetilde{\Psi}^{(R_j)}_P(q) \big|_{q \searrow e^{2\pi i\frac{s}{r}}}$, so the non-perturbative part of $Z_{M_3}(K)$ can be expressed as
\begin{align}
- g(Pr;s) \sum_{j=0}^{3} \frac{1}{2}\sqrt{\frac{r/s}{i}} \sum_{b=1}^{P-1} \sqrt{\frac{2}{P}} \sin \frac{\pi R_j b}{P} \sum_{n=1}^{2Ps-1} \psi_{2P}^{(b)}(n) e^{-2 \pi i \frac{1}{4P} \frac{r}{s} n^2} \Big( 1 - \frac{n}{Ps} \Big)	\label{rat-nonpert2}
\end{align} 
By using \eqref{sine-id}, \eqref{rat-nonpert2} becomes
\begin{align}
- 2 g(Pr;s) \sqrt{\frac{r/s}{i}} \sqrt{\frac{2}{P}} \sum_{b=1}^{P-1} \prod_{j=1}^3 (-1)^b \sin \frac{\pi b}{P_j} \sum_{n=1}^{2Ps-1} \psi_{2P}^{(b)}(n) e^{-2 \pi i \frac{1}{4P} \frac{r}{s} n^2} \Big( 1 - \frac{n}{Ps} \Big)	\,	.	\label{rat-nonpert3}
\end{align}
$b$'s such that $\sin \frac{\pi b}{P_j}$ is nonzero are not divisible by any of $P_j$'s.
For such $b$'s in $[ 1,P-1 ]$, $n$'s that give a nonzero result are $n= \pm b$ mod $2P$.
Therefore, \eqref{rat-nonpert3} can be written as
\begin{align}
- 4 g(Pr;s) \sqrt{\frac{r/s}{i}} \sqrt{\frac{2}{P}} \sum_{n=1}^{Ps-1} (-1)^n  e^{-2 \pi i \frac{1}{4P} \frac{r}{s} n^2} \Big( 1 - \frac{n}{Ps} \Big) \prod_{j=1}^3 \sin \frac{\pi n}{P_j}	\,	,	\label{rat-nonpert4}
\end{align}
which agrees with the contribution from the non-abelian flat connections obtained via resurgent analysis for $Z_{M_3}(K) \simeq g(Pr;s) \sum_{j=0}^3 \widetilde{\Psi}^{(R_j)}_P(q) \big|_{q \searrow e^{2\pi i\frac{s}{r}}}$.


\subsubsection*{Example: $H=1$}

We provide an example for $H=1$, when $K$ is taken to be a rational number.
We discuss an example for $H \geq 2$ in Appendix \ref{app:h2g}.
For concreteness, we consider the Poincar\'e sphere $\Sigma(2,3,5)$ where $(P_1,P_2,P_3)=(2,3,5)$.

When $K$ is an integer, there two sets of poles $y=-n \pi i$ in \eqref{resurg-CS-f3} with 
\begin{align}
n=
\begin{cases}
1, 11, 19, 29, 31, 41, 49, 59	\\
7, 13, 17, 23, 37, 43, 47, 53
\end{cases}	\quad	\text{mod } 60
\end{align}
and their total sum of the residues after regularization are
\begin{align}
8 \pi \sqrt{3 \bigg(\frac{5}{8}-\frac{\sqrt{5}}{8}\bigg)} e^{-2 \pi i K \frac{1}{120}}	\,	,	\qquad
8 \pi \sqrt{3 \bigg(\frac{5}{8}+\frac{\sqrt{5}}{8}\bigg)} e^{-2 \pi i K \frac{49}{120}}
\label{csinv-int235}
\end{align}
respectively.
They correspond to two non-abelian flat connections. 
Their Chern-Simons invariants
\begin{align}
-\frac{n^2}{4P}	\quad	\text{mod } 1
\end{align}
can be read off from the factor $e^{-2 \pi i \frac{n^2}{4P}}$ in \eqref{res-1}, which are in this case 
\begin{align}
CS(\alpha_1) = -\frac{1}{120}	\,	,	\qquad
CS(\alpha_2) = -\frac{49}{120}	\,	,	\qquad	\text{mod } 1	\,	,
\end{align}
respectively.	\\

When $K= \frac{r}{s}$, a natural way to sort the poles would be to consider sets of poles that have the same factor $e^{-2\pi i \frac{r}{s} \frac{n^2}{4P}}$, \textit{i.e.} the same $-\frac{r}{s} \frac{n^2}{4P}$ mod 1. 
If we denote $N$ as the number of sets of poles for the case of the integer level $K=r$, for examples that we work where $s$ is an odd integer, the number of sets of poles for the case of $K=\frac{r}{s}$ is given by $\frac{s+1}{2}N$.
For example, when $s=7$, poles at $y=-n\pi i$ in \eqref{irsr-int} can be classified by 8 sets of poles
\begin{empheq}[left = { n = \empheqlbrace}]{alignat*=3}
&    
\left\{
\begin{array}{l l l}
\alpha_{1,1} :	&	1,29,41,71,139,169,181,209,211,239,251,281,349,379,391,419		\\
\alpha_{1,2} :	&	11,31,59,101,109,151,179,199,221,241,269,311,319,361,389,409	\\
\alpha_{1,3} :	&	19,61,79,89,121,131,149,191,229,271,289,299,331,341,359,401	\\
\alpha_{1,4} :	&	49,91,119,161,259,301,329,371	
\end{array}
\text{ mod } 420
\right\}
&&
\leftarrow \alpha_1
\\
&
\left\{
\begin{array}{l l}
\alpha_{2,1} :	&	7,77,133,203,217,287,343,413	\\
\alpha_{2,2} :	&	13,43,83,97,113,127,167,197,223,253,293,307,323,337,377,407	\\
\alpha_{2,3} :	&	17,53,67,73,137,143,157,193,227,263,277,283,347,353,367,403	\\
\alpha_{2,4} :	&	23,37,47,103,107,163,173,187,233,247,257,313,317,373,383,397	
\end{array}
\text{ mod } 420
\right\}
&&
\leftarrow \alpha_2
\end{empheq}
and the residues for $(\alpha_{1,l})_{l=1,2,3,4}$ and $(\alpha_{2,l})_{l=1,2,3,4}$ are given by
\begin{align}
&8 \pi g(30r,7) \sqrt{3 \bigg(\frac{5}{8}-\frac{\sqrt{5}}{8}\bigg)} e^{-2 \pi i \frac{r}{7} \frac{1}{120}} \, (e^{2 \pi i \frac{r}{7} \cdot 0},-e^{2 \pi i\frac{r}{7} \cdot 6},2 e^{2 \pi i \frac{r}{7} \cdot 4},-e^{2 \pi i \frac{r}{7} \cdot 1} )	\label{csinv-rat235s7-1}	\\
&8 \pi g(30r,7) \sqrt{3 \bigg(\frac{5}{8}+\frac{\sqrt{5}}{8}\bigg)} e^{-2 \pi i \frac{r}{7} \frac{49}{120}} \, ( e^{2 \pi i \frac{r}{7} \cdot 0},0,e^{2 \pi i \frac{r}{7} \cdot 5},-e^{2 \pi i  \frac{r}{7} \cdot 3} )	\label{csinv-rat235s7-2}
\end{align}
respectively.
We see that the exponents in \eqref{csinv-rat235s7-1} and \eqref{csinv-rat235s7-2} are given by $2\pi i\frac{r}{s} (CS(\alpha)+c)$ where $c$ is an integer, which would be all the same with $2\pi i r CS(\alpha)$ if $s$ were set to 1.

Transseries parameters $m_{(a_0,0)}^{\bbbeta}$ for $s=7$ and $r \in \mathbb{Z}$ associated to $y=-n \pi i$ are given by
\begin{empheq}[left={m_{(a_0,0)}^{\bbbeta} = \empheqlbrace}]{alignat*=2}
\text{for } \bbbeta=(\alpha_{1},-\frac{1}{420}n^2)	&	
\left\{		
\begin{array}{l l l }
\alpha_{1,1}	\hspace{1mm}
\left\{ 
\begin{array}{l l l }
+1	&\quad n = 1,29,71,139,181,209,251,379		\\
-1	&\quad n = 41,169,211,239,281,349,391,419	
\end{array}	\right.			\vspace{1mm}\\	
\alpha_{1,2}	\hspace{1mm}
\left\{ 
\begin{array}{l l l }
+1	&\quad n = 11,199,241,269,311,319,361,389	\\
-1	&\quad n = 31,59,101,109,151,179,221,409	
\end{array}	\right.			\vspace{1mm}\\
\alpha_{1,3}	\hspace{1mm}
\left\{ 
\begin{array}{l l l }
+1	&\quad n = 19,61,79,89,121,131,149,191		\\
-1	&\quad n = 229,271,289,299,331,341,359,401	
\end{array}	\right.			\vspace{1mm}\\
\alpha_{1,4}	\hspace{1mm}
\left\{ 
\begin{array}{l l l }
+1	&\quad n = 259,301,329,371	\\
-1	&\quad n = 49,91,119,161	
\end{array}	\right.			
\end{array}
\right.	\\
\text{for } \bbbeta=(\alpha_{2},-\frac{1}{420}n^2)	&	
\left\{			
\begin{array}{l l l }
\alpha_{2,1}	\hspace{1mm}
\left\{ 
\begin{array}{l l l }
+1	&\quad n = 7,77,133,203	\\
-1	&\quad n = 217,287,343,413	
\end{array}	\right.			\vspace{1mm}\\	
\alpha_{2,2}	\hspace{1mm}
\left\{ 
\begin{array}{l l l }
+1	&\quad n = 13,83,127,197,253,307,323,377	\\
-1	&\quad n = 43,97,113,167,223,293,337,407	
\end{array}	\right.			\vspace{1mm}\\
\alpha_{2,3}	\hspace{1mm}
\left\{ 
\begin{array}{l l l }
+1	&\quad n = 17,67,73,137,143,193,263,367	\\
-1	&\quad n = 53,157,227,277,283,347,353,403	
\end{array}	\right.			\vspace{1mm}\\
\alpha_{2,4}	\hspace{1mm}
\left\{ 
\begin{array}{l l l }
+1	&\quad n = 23,187,247,257,313,317,373,383	\\
-1	&\quad n = 37,47,103,107,163,173,233,397	
\end{array}	\right.			
\end{array}
\right.	
\end{empheq}
mod 420 where $Z_{\alpha_1} = 8 \pi g(30r,7) \sqrt{3 \Big(\frac{5}{8}-\frac{\sqrt{5}}{8}\Big)}$ and $Z_{\alpha_2}=8 \pi g(30r,7) \sqrt{3 \Big(\frac{5}{8}+\frac{\sqrt{5}}{8}\Big)}$.


\section{Resurgent analysis for torus knot complement in $S^3$}
\label{sec:torusknot}

A new $q$-series invariant with an additional variable $x$ for the knot complement $S^3 \backslash \mathcal{K}$ in $S^3$ was studied in \cite{Gukov-Manolescu}.
It is denoted as $F_{\mathcal{K}}(x; q)$, and we call it the homological block for a knot complement as $\widehat{Z}$ is called the homological block for the closed 3-manifolds.
One of conjectures in \cite{Gukov-Manolescu} states that a two variable invariants $f_{\mathcal{K}}(x;q)=\frac{F_{\mathcal{K}}(x;q)}{x^{\frac{1}{2}}-x^{-\frac{1}{2}}}$ with integer coefficients is obtained from the Borel resummation of the asymptotic expansion of the reduced colored Jones polynomial with a color ${R}$ around the trivial flat connection discussed in \cite{Rozansky-MM}, which is 
\begin{align}
\widetilde{J}_{\mathcal{K}}(R, e^\hbar) = \sum_{m=1}^\infty \frac{P_m(x)}{\Delta_{\mathcal{K}}(x)^{2m+1}} \hbar^m = \sum_{m=0}^{\infty} \sum_{j=0}^m c_{m,j} {R}^j \hbar^{m}
\end{align}
where $q=e^\hbar$ and $x=q^R$.
$P_m(x) \in \mathbb{Q}[x^{\pm1}]$ is a Laurent polynomial with $P_0(x)=1$ and $\Delta_{\mathcal{K}}(x)$ is the Alexander polynomial.
Here the coefficients $c_{m,j}$ are the Vassiliev invariants for a knot $\mathcal{K}$.
Therefore, $F_{\mathcal{K}}(x;q)$ can be regarded as an analytic continuation of the colored Jones polynomial.
In this section, we discuss aspects of the homological block $F_{\mathcal{K}}(x;q)$ for a knot complement $S^3\backslash \mathcal{K}$ as a full partition function of an analytically continued Chern-Simons theory by studying the torus knot complement in $S^3$ via resurgent analysis, which directly indicates that it is indeed an analytic continuation of the colored Jones polynomial. 	\\

The integral expression of the $G=SU(2)$ Chern-Simons partition function for a torus knot complement in $S^3$ was obtained in \cite{Lawrence-Rozansky, Beasley-Wilson}.\footnote{More generally, in \cite{Beasley-Wilson}, the integral expression of $SU(2)$ Chern-Simons partition function was obtained for Seifert loops, which contains the case of the torus knot in $S^3$.
In this paper, we discuss the case of the torus knot in $S^3$ for simplicity and the case of Seifert loops can be done similarly.}
The unnormalized Jones polynomial for (the mirror of) torus knot $\mathcal{K}_{P,Q}$ with an irreducible representation of a dimension $R$ is given by
\begin{align}
J_{\mathcal{K}_{P,Q}}(R,K) = \frac{1}{2i} \sqrt{\frac{2}{K}} q^{-\frac{1}{4} P Q (R^2-1)} \sum_{j=0}^{R-1} q^{\frac{1}{4} (2j-R+1)^2 P Q} q^{\frac{1}{2} (2j-R+1) P} \big( q^{\frac{1}{2} ( (2j-R+1)Q +1 )} - q^{-\frac{1}{2} ( (2j-R+1)Q +1 )} \big)	\,		\label{jones-torus}
\end{align}
where $q=e^{\frac{2 \pi i}{K}}$ with $K \in \mathbb{Z}$.
The Jones polynomial \eqref{jones-torus} can be expressed as
\begin{align}
J_{\mathcal{K}_{P,Q}}(R,K) = D \bigg( \int_{C'} dy \, h(u,y) -2 \pi i \sum_{m=1}^{2PQ-1} \text{Res} \bigg( \frac{h(u,y)}{1-e^{-2Ky}}, y=\pi i m \bigg) \bigg)	\label{int-jones-torus}
\end{align}
where
\begin{align}
h(u,y) = e^{-\frac{K}{2\pi i} \frac{1}{PQ} y^2} (e^{K u y} - e^{-K u y}) \frac{(e^{y/P} - e^{-y/P}) (e^{y/Q} - e^{-y/Q}) }{e^y - e^{-y}}
\end{align}
with
\begin{align}
u:=\frac{R}{K}
\end{align} 
and the contour $C'$ is a line that passes from $-(1+i)\infty$, the origin, and to $(1+i)\infty$ in the $y$-plane.
The overall factor $D$ is 
\begin{align}
D = -\frac{1}{4 \pi} q^{\frac{PQ}{4}-\frac{1}{4}\frac{P}{Q} -\frac{1}{4} \frac{Q}{P}} \frac{e^{\frac{\pi i}{4} \text{sign}PQ}}{\sqrt{|PQ|}} q^{-\frac{PQ}{4}R^2}	\,	.
\end{align}
Here what we mean by the unnormalized Jones polynomial is that a normalization for the Jones polynomial of the unknot is chosen as
\begin{align}
J_{\mathbf{0}_1}(R,K) = \sqrt{\frac{2}{K}} \sin \frac{\pi}{K}R	\,	.
\end{align}
The residue part in \eqref{int-jones-torus} is given by
\begin{align}
i R \sum_{m=1}^{2PQ-1} (-1)^m e^{\pi i K u m} e^{-\frac{\pi i}{2} K \frac{1}{PQ} m^2} \sin \frac{\pi m}{P} \sin \frac{\pi m}{Q}	
\end{align}
but we can see that this residue part vanishes.\footnote{See also \cite{Beasley-Wilson}.}	
Therefore, the Jones polynomial is obtained from the Gaussian integral part of \eqref{int-jones-torus},
\begin{align}
J_{\mathcal{K}_{P,Q}}(R,K) = D  \int_{C'} dy \, e^{-\frac{K}{2\pi i} \frac{1}{PQ} y^2} (e^{K u y} - e^{-K u y}) \frac{(e^{y/P} - e^{-y/P}) (e^{y/Q} - e^{-y/Q}) }{e^y - e^{-y}}	\, . 	\label{int-jones-torus2}
\end{align}
\\

Another type of residues arises by moving the contour from $C'$ to the stationary phase points, $y = \pm \pi i P Q u$, of $h(u,y)$ in \eqref{int-jones-torus2}.
Denoting such contours as $C^{'}_{\pm}$, the integral \eqref{int-jones-torus2} can be expressed as
\begin{align}
\begin{split}
\int_{C'} dy \, h(u,y) = 
&e^{\frac{\pi i}{2} K P Q u^2} \bigg( \int_{C'_{+}} dy \, h_+(u,y) - \int_{C'_{-}} dy \, h_-(u,y) \bigg)	\\
&+2 \pi i \, e^{\frac{\pi i}{2} K P Q u^2} \sum_{m=1}^{\lfloor PQu \rfloor} \bigg( \text{Res} \Big[ h_+(u,y)	\, , y=m \pi i \Big] -\text{Res} \Big[ h_-(u,y)	\, , y=-m \pi i \Big] \bigg)	\label{torus-gauss}
\end{split}
\end{align}
where $h_{\pm}(u,y) = e^{-\frac{K}{2\pi i} \frac{1}{PQ} (y \mp \pi i P Q u)^2} \frac{(e^{y/P} - e^{-y/P}) (e^{y/Q} - e^{-y/Q}) }{e^y - e^{-y}}$.
This can be simplified to
\begin{align}
\begin{split}
&2 e^{\frac{\pi i}{2} K P Q u^2}\int_{C'} dy \, e^{-\frac{K}{2\pi i} \frac{1}{PQ} y^2} \frac{(e^{\frac{1}{P}(y + \pi i P Q u)} - e^{-\frac{1}{P}(y + \pi i P Q u)}) (e^{\frac{1}{Q}(y + \pi i P Q u)} - e^{-\frac{1}{Q}(y + \pi i P Q u)}) }{e^{(y + \pi i P Q u)} - e^{-(y + \pi i P Q u)}}	\\
&-8 \pi i \sum_{m=1}^{\lfloor PQu \rfloor} (-1)^m e^{\pi i K u m} e^{-2 \pi i K \frac{m^2}{4PQ} } \sin \frac{\pi m}{P} \sin \frac{\pi m}{Q}	\,	.	\label{torus-fin}
\end{split}
\end{align}
The residue part of \eqref{torus-fin} vanishes if $u< \frac{1}{PQ}$.

The integral part of \eqref{torus-fin} corresponds to the contribution from the trivial (abelian) flat connection and the residue part of \eqref{torus-fin} is from non-abelian flat connections of the torus knot complement in $S^3$.	
For a $(P,Q)$ torus knot, there are $\frac{1}{2}(P-1)(Q-1)$ non-abelian flat connections \cite{Klassen-rep} and this can also be seen from the number of distinct Chern-Simons invariant $- \frac{1}{4 PQ}m^2$ mod 1, which can be read off from the factor $e^{-2 \pi i K \frac{m^2}{4PQ}}$ in the residue part of \eqref{torus-fin}.


\subsection*{Homological block for torus knot complement in $S^3$}

We would like to discuss that $F_{\mathcal{K}}(x;q)$ for the torus knot complement can be obtained from analytic continuations of $K$ and $R$ in the Gaussian integral part of \eqref{torus-gauss}.
For the calculation of the homological block for the torus knot complement,
we expand $\frac{(e^{y/P} - e^{-y/P}) (e^{y/Q} - e^{-y/Q}) }{e^y - e^{-y}}$ in \eqref{torus-gauss} as
\begin{align}
\frac{(e^{y/P} - e^{-y/P}) (e^{y/Q} - e^{-y/Q}) }{e^y - e^{-y}} = \sum_{n=0}^{\infty} \xi_{2PQ}(n) e^{-\frac{y}{PQ}n}	\label{expansion-sinh}
\end{align}
where 
\begin{align}
\xi_{2PQ}(n) = 
\begin{cases} 
+ 1	&	\text{if } n \equiv PQ+P+Q \ \text{or}	\ PQ-P-Q	\ \text{mod } 2PQ	\\
- 1	&	\text{if } n \equiv PQ-P+Q \ \text{or}	\ PQ+P-Q	\ \text{mod } 2PQ	\\
0	&	\text{otherwise}
\end{cases}	.
\end{align}
Here, we took $PQ>0$ and $\text{Re }y >0$.
After an analytic continuation of $K$ where we choose $\text{Im }K <0$ for convergence, we take the integral contour as $\gamma$.
Then the Gaussian integral part can be written as
\begin{align}
D \int_\gamma dy \,  e^{\frac{\pi i}{2} K PQ u^2} ( e^{-\frac{K}{2\pi i} \frac{1}{PQ} (y-\pi i PQ u)^2} - e^{-\frac{K}{2\pi i} \frac{1}{PQ} (y+\pi i PQ u)^2} ) \sum_{n=0}^{\infty} \xi_{2PQ}(n) e^{-\frac{y}{PQ}n}	\,	.	\label{int-torus}
\end{align}
This becomes
\begin{align}
-D e^{\frac{\pi i}{2} K PQ u^2} \sum_{n=0}^{\infty} \xi_{2PQ}(n) (e^{\pi i u n} - e^{-\pi i u n}) q^{\frac{n^2}{4PQ}} \int_{\gamma} dy \, e^{-\frac{K}{2\pi i} \frac{1}{PQ} (y+ \frac{\pi i}{K}n)^2}
\end{align}
where the last integral term gives the overall factor, $\pi i \big( \frac{2}{K} |PQ| \big)^{1/2} e^{-\frac{\pi i}{4} \text{sign}\big( \frac{1}{PQ} \big)}$.
Thus, we obtain
\begin{align}
F_{\mathcal{K}}(x;q) = -\frac{1}{2i} \frac{1}{\sqrt{2K}} q^{\frac{PQ}{4}-\frac{1}{4}\frac{P}{Q} -\frac{1}{4} \frac{Q}{P}} \sum_{n=0}^{\infty} \xi_{2PQ}(n) (x^{n/2} - x^{-n/2}) q^{\frac{n^2}{4PQ}}	\label{torus-hom-block}
\end{align}
where
\begin{align}
x=e^{2\pi i u}	\,	.
\end{align} 
This agrees with the homological block $F_{\mathcal{K}}(x,q)$ of the torus knot in \cite{Gukov-Manolescu} up to an overall factor.

Or we may take the integral part of \eqref{torus-fin} and take an expansion
\begin{align}
\frac{(e^{y/P} - e^{-y/P}) (e^{y/Q} - e^{-y/Q}) }{e^y - e^{-y}} = -\frac{1}{2} \sum_{n=0}^{\infty} \xi_{2PQ}(n) (e^{\frac{y}{PQ}n} - e^{-\frac{y}{PQ}n})	\,	,	\label{expansion-sinh2}
\end{align}
when we don't specify the range of $y$.
Then, by substituting \eqref{expansion-sinh2} into the integral \eqref{torus-fin} with analytic continuations of $K$ and $R$ and also with the integration contour $\gamma$, we have the same homological block \eqref{torus-hom-block}.


\subsection*{Resurgent analysis}

As in the case of the Seifert manifolds, the Borel resummation of the Borel transform of the perturbative expansion of the homological block $F_{\mathcal{K}}(x;q)$\footnote{While preparing the manuscript, we found that \cite{Fuji:2020ltq} appeared where resurgent analysis for the WRT invariant of Seifert loop was considered while $q$ is analytically continued but the color is fixed as a positive integer, $R \in \mathbb{Z}_+$, throughout.
Our discussion in this section is about resurgent analysis for the homological block $F_{\mathcal{K}}(x;q)$ with a general complex $R$, including the case of the specialization $F_{\mathcal{K}}(x=q^R;q)$ with a given $R\in \mathbb{Z}_+$.} 
is given by 
\begin{align}
\begin{split}
\int_\gamma dy \, e^{-\frac{K}{2\pi i} \frac{1}{PQ} y^2} ( e^{- K u y} - e^{ K u y} ) \frac{(e^{y/P} - e^{-y/P}) (e^{y/Q} - e^{-y/Q}) }{e^y - e^{-y}}	\,	.	\label{Bsum-torus}
\end{split}
\end{align}

Given the Borel resummation \eqref{Bsum-torus}, the contributions from all non-abelian flat connections can also be recovered.
For $K \in \mathbb{Z}_+$ and also for $R=K u \in \mathbb{Z}_+$, we deform the integral contour from $\gamma$ to $C'$.
We note that \eqref{Bsum-torus} doesn't pick poles along the process due to $e^{ K u y}-e^{- K u y} = e^{Ry}- e^{-Ry}$ factor in the numerator.
Then the \eqref{Bsum-torus} becomes \eqref{int-jones-torus2} and then by shifting the integral contours $C'$ to $C'_+$ and $C'_-$ for the part that contains $e^{-Kuy}$ and $e^{Kuy}$, respectively, the residues in \eqref{torus-gauss} are recovered.\footnote{We may consider resurgent analysis from
\begin{align}
2 \int_{\gamma} dy \, e^{-\frac{K}{2\pi i} \frac{1}{PQ} y^2} \frac{(e^{\frac{1}{P}(y + \pi i P Q u)} - e^{-\frac{1}{P}(y + \pi i P Q u)}) (e^{\frac{1}{Q}(y + \pi i P Q u)} - e^{-\frac{1}{Q}(y + \pi i P Q u)}) }{e^{(y + \pi i P Q u)} - e^{-(y + \pi i P Q u)}}		\,	,	\label{Bsum-torus2}
\end{align}
which agrees with the result of \cite{Gukov-Manolescu} for the trefoil knot when $(P,Q)=(2,3)$.
As discussed above, upon $K \in \mathbb{Z}_+$ and $R=K u \in \mathbb{Z}_+$, we deform $\gamma$ to $C'$ as in Figure \ref{resum3}.
Then, we can see that \eqref{int-jones-torus2} or \eqref{torus-fin} is obtained from \eqref{Bsum-torus2}.
}
Therefore, contributions from all non-abelian flat connections can be recovered from the homological block $F_{\mathcal{K}}(x;q)$.

Thus, the homological block $F_{\mathcal{K}}(x;q)$ for a torus knot, which is obtained by analytic continuations of $K$ and $R$ in the integral expression \eqref{torus-gauss} that gives the Jones polynomial when $K$ is an integer, contains the information of all flat connections so it can be regarded as a full partition function of an analytically continued Chern-Simons theory. 
This indicates that the homological block $F_{\mathcal{K}}(x;q)$ is indeed an analytic continuation of the colored Jones polynomial which is a full $G=SU(2)$ partition function with an integer level $K$.
We expect that this argument hold for general knot complements in 3-manifolds.
If this is so, it implies that the surgery operation discussed in \cite{Gukov-Manolescu} is indeed a full topological quantum field theoretical operation with no flat connections left behind.	\\

As examples, we provide the transseries for the case of the trefoil knot $(P,Q)=(2,3)$ for several $u$.
\begin{empheq}[left = {}]{alignat*=5}
&
u = \left\{
\begin{array}{l}	\vspace{0.3mm}
\frac{1}{6}	\\	\vspace{0.3mm}
\frac{4}{5}	\\	\vspace{0.3mm}
\frac{5}{6}	\\	\vspace{0.3mm}
\frac{7}{6}	\\	\vspace{0.3mm}
\frac{11}{7}	\\	\vspace{0.3mm}
\frac{11}{6}	\\	\vspace{0.3mm}
\frac{17}{9}	\\	\vspace{0.3mm}
\frac{13}{6}
\end{array}
\right.
\quad
&&
\leadsto
&&	\quad
\text{poles} = 
\left\{
\begin{array}{l}
1	\\
1	\\
1,5	\\
1,5,7	\\
1,5,7	\\
1,5,7,11	\\
1,5,7,11	\\
1,5,7,11,13	
\end{array}
\right.
&&
\leadsto
&&	\quad
4 \pi i \sqrt{3}  e^{\frac{23 \pi i K}{12}} 
\left\{
\begin{array}{l}
e^{\frac{\pi i K }{6}}	\\	
e^{\frac{4 i \pi  K}{5}}	\\
e^{\frac{5 \pi i K}{6}},-e^{\frac{\pi i K }{6}}	\\	
e^{\frac{7 \pi i K }{6}},-e^{\frac{11\pi i K }{6}},-e^{\frac{\pi i K }{6}}	\\
e^{\frac{11 i \pi  K}{7}},-e^{\frac{13 i \pi  K}{7}},-e^{i \pi  K}	\\	
e^{\frac{11 \pi i K}{6}},-e^{\frac{7 \pi i K }{6}},-e^{\frac{5 \pi i K }{6}},e^{\frac{\pi i K}{6}}	\\	
e^{\frac{17 i \pi  K}{9}},-e^{\frac{13 i \pi  K}{9}},-e^{\frac{11 i \pi  K}{9}},e^{\frac{7 i \pi  K}{9}}	\\
e^{\frac{\pi i K }{6}},-e^{\frac{5 \pi i K }{6}},-e^{\frac{7 \pi i K }{6}},e^{\frac{11 \pi i K }{6}},e^{\frac{\pi i K}{6}}	
\end{array}
\right.	\,	.
\end{empheq}

As a remark, if taking $x= e^{2 \pi i \frac{R}{K}}=q^R$ in the homological block \eqref{torus-hom-block} where $R$ is an integer $R \in \mathbb{Z}_{+}$ and $K$ is still analytically continued, \eqref{torus-hom-block} can be simplified to
\begin{align}
F_{\mathcal{K}}(q^R;q) = \frac{1}{2i} \sqrt{\frac{2}{K}} q^{\frac{1}{4} ( PQ - \frac{P}{Q} - \frac{Q}{P} )} \sum_{n=0}^{PQR} \xi_{2PQ}(n) \, q^{- R n/2} \, q^{\frac{n^2}{4PQ}}	\,	.	\label{torus-hb-lim2}
\end{align}
This agrees with the Jones polynomial \eqref{jones-torus} with an analytically continued $q$.
So the specialization $F_{\mathcal{K}}(x=q^R;q)$ with $R \in \mathbb{Z}_+$ of the homological block for torus knot agrees with an analytic continuation of $q$ in the colored Jones polynomial with a given representation $R$, which we also expect so for a general knot.\footnote{We found that this is also observed in \cite{Park:2020edg}.}	\\

As another remark, in the previous discussion for the Seifert manifolds, we saw that the resurgent analysis provide a nearly modular property for the homological block.
For torus knot complement, we have
\begin{align}
F_{\mathcal{K}}(x=e^{2\pi i u};q=e^{\frac{2 \pi i}{K}}) = -8 \pi i \sum_{m=1}^{\lfloor PQu \rfloor} (-1)^m e^{\pi i K u m} e^{-2 \pi i K \frac{m^2}{4PQ} } \sin \frac{\pi m}{P} \sin \frac{\pi m}{Q} + \text{pert.}		\label{fk-full}
\end{align}
for $K, R \in \mathbb{Z}_+$ where $u=\frac{R}{K}$. 
The perturbative part is obtained by taking the Taylor expansion of the rational function of sine hyperbolic function in \eqref{torus-fin} around $y=0$ and evaluating the Gaussian integral term by term with the factor $e^{-\frac{K}{2\pi i} \frac{1}{PQ} y^2}$, which is \cite{Rozansky-MM}
\begin{align}
\frac{1}{e^{\hbar/2}-e^{-\hbar/2}} e^{\frac{\hbar}{4} (PQ-\frac{P}{Q}-\frac{Q}{P})} \sum_{m=0}^{\infty} \frac{1}{m!} \Big( \frac{\hbar}{4PQ} \Big)^m \Big( \frac{\partial}{\partial y} \Big)^{2m} \frac{ x^{\frac{1}{2}}e^y-x^{-\frac{1}{2}}e^{-y} }{\Delta_{\mathcal{K}_{P,Q}}(x^{\frac{1}{2}}e^y)} \bigg|_{y=0}	\label{fk-pert}
\end{align}
where
\begin{align}
\Delta_{\mathcal{K}_{P,Q}}(x) = \frac{(x -x^{-1}) (x^{PQ}-x^{-PQ})}{(x^{P}-x^{-P})(x^{Q}-x^{-Q})}	\,	.	\label{fk-alexander}
\end{align}
For some closed 3-manifolds $M_3$, the homological block for $M_3$ is known to have near modularity, which may be understood in the context of the 3d-3d correspondence.
From the perspective of the 3d $\mathcal{N}=2$ theory $T[M_3,SU(2)]$, the homological block $\widehat{Z}(q)$ of closed 3-manifolds is given by the $D^2 \times_q S^1$ partition function of $T[M_3,SU(2)]$.
Since $D^2 \times_q S^1 \simeq \mathbb{R}_+ \times T^2$, if there is no $\mathbb{R}_+$, the partition function would be modular, but due to the bulk part of $T[M_3]$, the modularity would be spoiled in such a way that the partition function of $T[M_3]$ is nearly modular.
As in the case that $M_3$ is a Seifert manifold, it is expected that \eqref{fk-full}\footnote{
By introducing a periodic function $\varphi^{(a)}_{2P}$
\begin{align}
\varphi_{2P}^{(a)}(n) = 
\begin{cases} 
1	&	n = \pm a \, \text{ mod } 2P	\\ 
0	&	\text{otherwise} 
\end{cases}	\,	,
\end{align}
we denote $F_{\mathcal{K}}^{(a)}(x;q)$ as
\begin{align}
F_{\mathcal{K}}^{(a)}(x;q) = \sum_{n=0}^{\infty} \varphi^{(a)}_{2PQ}(n) (x^{n/2}-x^{-n/2}) q^{\frac{n^2}{4PQ}}	\,	.
\end{align}
Then, $F_{\mathcal{K}}(x;q)$ can be expressed as 
\begin{align}
F_{\mathcal{K}}(x;q) = F_{\mathcal{K}}^{(PQ-P-Q)}(x;q) - F_{\mathcal{K}}^{(PQ-P-Q)}(x;q)	
\end{align}
where 
\begin{eqnarray}
\begin{split}
F^{(a)}_{\mathcal{K}}(x=e^{2\pi i u};q=e^{\frac{2 \pi i}{K}}) &= 4 \pi i \sum_{m=1}^{\lfloor PQu \rfloor}  (-1)^m e^{\pi i K u m} e^{-2 \pi i K \frac{m^2}{4PQ} } \cos \Big( \frac{\pi m a}{PQ} \Big)	\\ 
&+ \frac{1}{e^{\hbar/2}-e^{-\hbar/2}} e^{\frac{\hbar}{4} (PQ-\frac{P}{Q}-\frac{Q}{P})} \sum_{m=0}^{\infty} \frac{1}{m!} \Big( \frac{\hbar}{4PQ} \Big)^m \Big( \frac{\partial}{\partial y} \Big)^{2m} \frac{ (x^{\frac{1}{2}}e^y)^{PQ-a} + (x^{-\frac{1}{2}}e^{-y})^{PQ-a} }{ (x^{\frac{1}{2}}e^y)^{PQ} + (x^{-\frac{1}{2}}e^{-y})^{PQ} } \bigg|_{y=0}	\,	.	\label{fk-part}
\end{split}
\end{eqnarray}
\eqref{fk-part} might be a more appropriate candidate for studying the near modularity.
}
exhibits a nearly modular property of $F_{\mathcal{K}}(x;q)$. 
It will be interesting to study further the modularity of $F_{\mathcal{K}}(x;q)$.


\subsection*{Other roots of unity}

We may also consider $F_{\mathcal{K}}(x;q)$ at other roots of unity.
The unnormalized Jones polynomial for the torus knot can be expressed as \cite{Lawrence-Rozansky}
\begin{align}
\hspace{-5mm}\frac{1}{4 i K \sqrt{|PQ|}} z^{PQ\big( PQ - \frac{P}{Q} -\frac{Q}{P} \big)} z^{-(PQ R)^2} e^{\frac{\pi}{4} \text{sign }PQ} 
\sum_{\substack{\beta=1 \\ K \nmid \beta}}^{2PQK-1} z^{-\beta^2} \frac{z^{2PQR\beta}-z^{-2PQR\beta}}{z^{2PQ\beta} - z^{-2PQ\beta}} (z^{2Q\beta}-z^{-2Q\beta}) (z^{2P\beta}-z^{-2P\beta})	\label{jones-sum}
\end{align}
where $z=e^{\frac{\pi i}{2KPQ}}$.
As in the case of the Seifert manifolds, considering the Galois action on $z$, $z$ is replaced with another primitive $4PQK$-th root of unity, $z=e^{\frac{s \pi i}{2KPQ}}$ where $s$ is coprime to $4PQK$.
With such an expression, the integral expression can be obtained as in the case of an integer $K$ and up to an overall factor\footnote{There are additional overall factors that depend on $s$ which arise from $i$, $K$, and $e^{\frac{\pi}{4} \text{sign }PQ}$ in the overall factor of \eqref{jones-sum} after the Galois action, which we will omit.} it is given by
\begin{align}
J_{\mathcal{K}_{P,Q}}(R, q=e^{2\pi i\frac{s}{r}}) \simeq D \bigg( \sum_{t=0}^{s-1} \int_{C'} dy \, h(u,y) e^{-2Kty} -2 \pi i \sum_{m=1}^{2PQs-1} \text{Res} \bigg( \frac{h(u,y)}{1-e^{-2Ky}}, y=\pi i m \bigg) \bigg)		\label{int-jones-torus-rat}
\end{align}
where $K= \frac{r}{s}$.
We can also see that the residue term in \eqref{int-jones-torus-rat} vanishes, so \eqref{int-jones-torus-rat} becomes
\begin{align}
J_{\mathcal{K}_{P,Q}} (R, q=e^{2\pi i\frac{s}{r}}) \simeq D \sum_{t=0}^{s-1} \int_{C'} dy \, e^{-\frac{K}{2\pi i} \frac{1}{PQ} y^2} (e^{K u y} - e^{-K u y}) \frac{(e^{y/P} - e^{-y/P}) (e^{y/Q} - e^{-y/Q}) }{e^y - e^{-y}} e^{-2Kty}	\,	.	\label{int-jones-torus-rat2}
\end{align}

There is also another type of the residue that arises from moving the contour $C'$ to the contour that passes the saddle points, $y=\pm \pi i PQ u -2 \pi i P Q t$ for each $t$.
After some calculations, \eqref{int-jones-torus-rat2} can be expressed as
\begin{align}
\begin{split}
D g(PQR,s) \bigg( 
&2 e^{\frac{\pi i}{2} K P Q u^2}\int_{C'} dy \, e^{-\frac{K}{2\pi i} \frac{1}{PQ} y^2} \frac{(e^{\frac{1}{P}(y + \pi i P Q u)} - e^{-\frac{1}{P}(y + \pi i P Q u)}) (e^{\frac{1}{Q}(y + \pi i P Q u)} - e^{-\frac{1}{Q}(y + \pi i P Q u)}) }{e^{(y + \pi i P Q u)} - e^{-(y + \pi i P Q u)}}	\\
&-8 \pi i \sum_{m=1}^{\lfloor PQu \rfloor} (-1)^m e^{\pi i K u m} e^{-2 \pi i K \frac{m^2}{4PQ} } \sin \frac{\pi m}{P} \sin \frac{\pi m}{Q}	
 \bigg)
\end{split}	\label{jones-rat-full}
\end{align}
where $K=\frac{r}{s}$.	\\

In order to obtain the Chern-Simons partition function at other roots of unity in terms of the homological block, we do a similar calculation as in the case of an integer $K$, and we have
\begin{align}
F_{\mathcal{K}_{P,Q}} (q^R; q=e^{2\pi i\frac{s}{r}}) \simeq \frac{1}{2i \sqrt{2 r}} q^{\frac{1}{4} \big( PQ - \frac{P}{Q} - \frac{Q}{P} \big)} g(PQR,s) \sum_{n=0}^\infty \xi_{2PQ}(n) (x^{n/2} - x^{-n/2}) q^{\frac{n^2}{4PQ}} \Bigg|_{\stackrel{\hspace{-5mm}q \searrow e^{2 \pi i \frac{s}{r}}}{ \hspace{1mm} x = q^{R} \, , \, R \in \mathbb{Z}_+} }	\,	.	\label{hblock-tor-rat}
\end{align}
We can also see that \eqref{jones-rat-full} can be recovered from \eqref{hblock-tor-rat} via the Stokes phenomena.


\subsection*{Limit $q \searrow e^{\frac{2\pi i}{K}}$ with arbitrary $x$}

We may also consider a limit $q \searrow e^{\frac{2\pi i}{K}}$ with $R$ arbitrary in $x=e^{2 \pi i u} = e^{2\pi i \frac{R}{K}}$ in resurgence.
Since $R$ is not taken to be an integer, there are additional contributions by moving the contour from $\gamma$ to $C'_0$ as in Figure \ref{resum3}, 
\begin{align}
4 \pi i \sum_{m=1}^{\infty} e^{-\frac{\pi i}{2}K \frac{1}{PQ}m^2} (e^{\pi i R m}-e^{-\pi i R m}) (-1)^m \sin \frac{\pi m}{P} \sin \frac{\pi m}{Q}	\,	,	\label{torus-res-gen1}
\end{align}
which was originally zero when $R$ was set to an integer.
Also, by shifting the contour $C_0'$ to the contours $C_{\pm}'$ that pass $y = \pm \pi i PQ u$ for each integrand containing $e^{\pm Kuy}$, respectively, there is also a contribution,
\begin{align}
-8 \pi i \sum_{m=1}^{\lfloor PQ(\text{Re } u + \text{Im } u) \rfloor} (-1)^m e^{\pi i K u m} e^{-2 \pi i K \frac{m^2}{4PQ} } \sin \frac{\pi m}{P} \sin \frac{\pi m}{Q}	\,	.	\label{torus-res-gen2}
\end{align}
Therefore, the total residue is given by the sum of \eqref{torus-res-gen1} and \eqref{torus-res-gen2}, so
\begin{align}
\begin{split}
F_{\mathcal{K}}(x,q) \Big|_{q \searrow e^{\frac{2\pi i}{K}}} = & \, 4 \pi i \sum_{m=1}^{\infty} (-1)^m e^{-\frac{\pi i}{2}K \frac{1}{PQ}m^2} (e^{\pi i R m}-e^{-\pi i R m}) \sin \frac{\pi m}{P} \sin \frac{\pi m}{Q} 	\\
&-8 \pi i \sum_{m=1}^{\lfloor PQ(\text{Re } u + \text{Im } u) \rfloor} (-1)^m e^{- \frac{\pi i}{2} K \frac{1}{PQ}m^2 } e^{\pi i R m} \sin \frac{\pi m}{P} \sin \frac{\pi m}{Q}
+\text{pert.}	\label{fk-lim2-full}
\end{split}
\end{align}
where the perturbative part is given by the integral part of \eqref{torus-fin}, which can be expressed as \eqref{fk-pert} with $\hbar = \frac{2\pi i}{K}$.
We don't have a nicer or closed form expression of \eqref{fk-lim2-full}, but we expect that \eqref{fk-lim2-full} provides a near modularity of $F_{\mathcal{K}}(x;q)$ for a general $x$. 
Also, this would be related to the asymptotic expansion of the Akutsu-Deguchi-Ohtsuki knot invariants, which was recently studied in the context of homological block $F_{\mathcal{K}}(x;q)$ \cite{Gukov:2020lqm}.

Similarly, we may also consider a limit to other roots of unity, $q \searrow e^{2\pi i \frac{s}{r}}$.
In this case, the limit is the same as \eqref{fk-lim2-full} with $K$ replaced by $\frac{r}{s}$ and with an overall factor $g(PQR,s)$.


\acknowledgments{I would like to thank Sergei Gukov for helpful discussion and remarks on the draft.
I am also grateful to Sungjay Lee for hospitality at the Korea Institute for Advanced Study (KIAS).
}


\begin{appendices}


\section{Transseries for the case of Seifert rational homology sphere}
\label{app:h2g}

In this appendix, we consider some aspects of transseries of an analytically continued Chern-Simons partition function on the Seifert rational homology spheres.


\subsection{Transseries for a general level $K$}
\label{app:trgenk}

In section \ref{ssec:rsgh2}, assuming the structure \eqref{csab}, we had an expression for an analytically continued Chern-Simons partition function \eqref{ibrs} as a Borel resummation, which we write here again
\begin{align}
\begin{split}
Z_{M_3}(K) &= e^{2\pi i K S_{\mathbb{0}}} \int_{\gamma} dy e^{-\frac{K}{2\pi i} \frac{H}{P}y^2} \frac{\prod_{j=1}^3 e^{y/P_j}-e^{-y/P_j}}{e^{y}-e^{-y}}	\\
&+\sum_{t=1}^{\frac{H-1}{2}} e^{2\pi i K S_{\mathbb{t}}} \int_{\gamma} dy \Big( e^{-\frac{K}{2\pi i} \frac{H}{P}(y+2\pi i \frac{P}{H}t)^2} +e^{-\frac{K}{2\pi i} \frac{H}{P}(y+2\pi i \frac{P}{H}(H-t))^2} \Big) \frac{\prod_{j=1}^3 e^{y/P_j}-e^{-y/P_j}}{e^{y}-e^{-y}}	
\end{split}	\label{aibrs}
\end{align}
when $H$ is odd where $CS(t) = \frac{P}{H} t^2$ mod 1 and $S_\mathbb{t} = CS(t) + \mathbb{Z}$.
When $H$ is even, we have
\begin{align}
\begin{split}
\hspace{-3mm}Z_{M_3}(K) &= e^{2\pi i K S_{\mathbb{0}}} \int_{\gamma} dy e^{-\frac{K}{2\pi i} \frac{H}{P}y^2} \frac{\prod_{j=1}^3 e^{y/P_j}-e^{-y/P_j}}{e^{y}-e^{-y}} 
+ e^{2\pi i K S_{\mathbb{\frac{H}{2}}}} \int_{\gamma} dy e^{-\frac{K}{2\pi i} \frac{H}{P}(y+ \pi i P)^2}  \frac{\prod_{j=1}^3 e^{y/P_j}-e^{-y/P_j}}{e^{y}-e^{-y}}	\\
&+\sum_{t=1}^{\frac{H}{2}-1} e^{2\pi i K S_{\mathbb{t}}} \int_{\gamma} dy \Big( e^{-\frac{K}{2\pi i} \frac{H}{P}(y+2\pi i \frac{P}{H}t)^2} +e^{-\frac{K}{2\pi i} \frac{H}{P}(y+2\pi i \frac{P}{H}(H-t))^2} \Big) \frac{\prod_{j=1}^3 e^{y/P_j}-e^{-y/P_j}}{e^{y}-e^{-y}}		\,	.
\end{split}	\label{aibrse}
\end{align}
We note that the sum of two exponential terms $e^{-\frac{K}{2\pi i} \frac{H}{P}(y+2\pi i \frac{P}{H}t)^2} +e^{-\frac{K}{2\pi i} \frac{H}{P}(y+2\pi i \frac{P}{H}(H-t))^2}$ in \eqref{aibrs} or \eqref{aibrse} originate from the sum of terms $(e^{-2\pi i \frac{t}{H} l} + e^{2\pi i \frac{t}{H} l})$ in \eqref{zt}, and they are related by the Weyl action.

When $\theta$ of $K= |K| e^{i \theta}$ is taken to be zero, the integral contour $\gamma$ in \eqref{aibrs} and \eqref{aibrse} should be changed to $C'_{t}$ and $C'_{H-t}$ as discussed in section \ref{ssec:rsgh2}.
Upon the change of the integration contour, the part $t \neq 0$ or $\frac{H}{2}$ is given by
\begin{align}
\begin{split}
&\hspace{-10mm}e^{2 \pi i K \frac{P}{H}t^2} \bigg( \int_{C'_t} dy \, e^{-\frac{K}{2\pi i} \frac{H}{P} \big( y + 2 \pi i \frac{P}{H} t \big)^2} \frac{\prod_{j=1}^3 (e^{\frac{1}{P_j}y} - e^{-\frac{1}{P_j}y})}{e^{y} - e^{-y}} + \int_{C'_{H-t}} dy \, e^{-\frac{K}{2\pi i} \frac{H}{P} \big( y + 2 \pi i \frac{P}{H} (H-t) \big)^2} \frac{\prod_{j=1}^3 (e^{\frac{1}{P_j}y} - e^{-\frac{1}{P_j}y})}{e^{y} - e^{-y}} \bigg)
\\
&\hspace{-12mm}+ 8 \pi e^{2 \pi i K \frac{P}{H}t^2} \bigg( \sum_{m=-\infty}^{-\lfloor \frac{2Pt}{H} \rfloor -1} e^{-\frac{\pi i}{2} K \frac{H}{P} \big( m + \frac{2Pt}{H} \big)^2} (-1)^m \prod_{j=1}^3 \sin \frac{\pi m}{P_j}	+\sum_{m=-\infty}^{-\lfloor \frac{2P(H-t)}{H} \rfloor-1} e^{-\frac{\pi i}{2} K \frac{H}{P} \big( m + \frac{2P(H-t)}{H} \big)^2} (-1)^m \prod_{j=1}^3 \sin \frac{\pi m}{P_j}	\bigg)	\,	.
\end{split}	\label{ratstk}
\end{align}
from \eqref{stokeh-1} and \eqref{stokeh-2} where $K$ is not necessarily an integer here and we chose a lift such that $S_{\mathbb{t}}=\frac{P}{H}t^2$.
For $m \in (-\infty, -\lfloor \frac{2P(H-t)}{H} \rfloor-1]$, the residue for a given pole $y=m \pi i$ is given by the sum of two exponential terms
\begin{align}
\begin{split}
&8\pi e^{2\pi i K \frac{P}{H}t^2} \Big( e^{-\frac{\pi i}{2}K \frac{H}{P}(m+\frac{2Pt}{H})^2} + e^{-\frac{\pi i}{2}K \frac{H}{P}(m+\frac{2P(H-t)}{H})^2} \Big) (-1)^m \prod_{j=1}^3 \sin \frac{\pi m}{P_j}	\\
&=8\pi \Big( e^{-\frac{\pi i}{2}K \frac{H}{P}m^2 -2 \pi i K t m} + e^{-\frac{\pi i}{2}K \frac{H}{P}m^2 +2 \pi i K ( (t-H)m+P(2t-H))} \Big) (-1)^m \prod_{j=1}^3 \sin \frac{\pi m}{P_j}	\,	.	\label{nabcont}
\end{split}
\end{align}

Considering an expected general structure \eqref{csab}, when $\theta$ is taken to be zero, upon the change of the contour, the analytically continued Chern-Simons partition function \eqref{csab} becomes \eqref{spt0}.
The case of $H=1$ fits well in the structure \eqref{csab} and \eqref{spt0}, and a pole $y=m\pi i$ corresponds to a lift $\bbbeta$ of a non-abelian flat connection $\beta$.
However, \eqref{nabcont} doesn't fit well in \eqref{spt0} as in the case of $H=1$, if a pole $y=m\pi i$ corresponds to a lift $\bbbeta$ of a non-abelian flat connection $\beta$ when $m \in (-\infty, -\lfloor \frac{2P(H-t)}{H} \rfloor-1]$.

More explicitly, consider a set of infinite number of poles $y=m \pi i$, $m=\{m_1, m_2, m_3, \ldots \}$, which give the same Chern-Simons invariant when $K$ is taken to be an integer, which would mean that they correspond to the same non-abelian flat connection $\beta$ when $K$ is an integer.
When $t=0$, a pole $y=m_j \pi i$ corresponding to a lift $\bbbeta_j$ gives a singe exponential factor in the residue and such a residue at $y=m_j \pi i$ gives $\frac{1}{2} m_{\mathbb{0}}^{\bbbeta_j} I_{\bbbeta_j}$ in \eqref{spt0}.

However, \eqref{nabcont} indicates that for a general $K$ a parameter $t$ gives two lifts given a pole $y=m_j \pi i$, say $\bbbeta_j^{(+)}$ and $\bbbeta_j^{(-)}$, which correspond to terms containing $e^{-\frac{\pi i}{2} K \frac{H}{P}m_j^2 - 2 \pi i K t m_j}$ and $e^{-\frac{\pi i}{2}K \frac{H}{P}m_j^2 +2 \pi i K ( (t-H)m_j+P(2t-H))}$, respectively.
Therefore, apparently \eqref{nabcont} indicates that there are two distinct lifts for a given pole $y=m \pi i$ and such two lifts $\bbbeta^{(+)}$ and $\bbbeta^{(-)}$ of an non-abelian flat connection are attached to an abelian flat connection $t$.
So the case of $H \geq 2$ is different from the case of $H=1$.	\\

In order to explain the case of rational homology spheres, we may speculate as follows.
Firstly, given \eqref{csab} and \eqref{spt0} as in \eqref{aibrse}, in the residue \eqref{ratstk} we see that given a $t$, certain poles don't contribute, give one exponential factor, or give two exponential factors, where the maximum number of exponential factors is the order of the Weyl orbit of the abelian flat connection $t$.
So depending on the value of $t$, given a pole $m_j$, it might be possible that there are at most two lifts of a non-abelian flat connection.
However, we don't have a good explanation why this is so.
For example, a lift of flat connections and the Weyl action are independent notion, so there is no good reason why for a certain range of $m$, the number of lifts associated to a given pole can be the order of the Weyl orbit of $t$.

Or, if a pole does correspond to a single lift, we may speculate that the analytically continued Chern-Simons partition function \eqref{csab} would be expressed as a sum over the elements in the Weyl orbits corresponding to abelian flat connections.\footnote{In resurgent analysis for the analytically continued Chern-Simons theory, a Lefschetz thimble is associated to a critical submanifold
\begin{align}
\text{Hom}(\pi_1(M_3), SL(2,\mathbb{C}))
\end{align}
rather than $\mathcal{M}_{\text{flat}}(M_3,SL(2,\mathbb{C})) = \text{Hom}(\pi_1(M_3), SL(2,\mathbb{C}))/SL(2,\mathbb{C})$ \cite{Gukov-Marino-Putrov} where the conjugation by $SL(2,\mathbb{C})$ contains the Weyl group action.
In particular, for the abelian flat connections, $\pi_0(\mathcal{M}^{\text{ab}}_{\text{flat}}(M_3,SL(2,\mathbb{C}))) = \text{Tor} \, H_1(M_3,\mathbb{Z})/\mathbb{Z}_2$ where $\mathbb{Z}_2$ is a Weyl group.
So it may be possible to consider a sum over $\text{Tor} \, H_1(M_3,\mathbb{Z})$.
}
Denoting $\tilde{a}$ as elements in the Weyl orbit corresponding to the abelian flat connection $a$ and $\tilde{\mathbb{a}}$ as lifts of $\tilde{a}$, it might be possible to have an expression
\begin{equation}
Z_{M_3}(K) = \sum_{\tilde{a} \in \text{Tor} \, H_1(M_3,\mathbb{Z})} e^{2 \pi i K S_{\mathbb{a}}} Z_{\tilde{a}}(K)	\label{acpartw}
\end{equation}
and
\begin{equation}
Z_{M_3}(K) = \sum_{\tilde{a} } \Big( I_{\tilde{\mathbb{a}}}(|K|,0) + \frac{1}{2} \sum_{\bbbeta \notin \text{abelian}} m_{\tilde{\mathbb{a}}}^{\bbbeta} I_{\bbbeta}(|K|,0)	 \Big)		\label{st0w}
\end{equation}
where
\begin{equation}
I_{\tilde{\mathbb{a}}}=e^{2\pi i K S_{\mathbb{a}}} Z_{\tilde{a}}	\,	,	\qquad	I_{\bbbeta}=e^{2\pi i K S_{\bbbeta} }Z_{\beta}	\,	.	\label{cabnab}
\end{equation}
Here, in this case, $\tilde{a}$ and $\widetilde{H-a}$, which are elements of the Weyl orbit corresponding to an abelian flat connection $a$, would have their own lifts $\tilde{\mathbb{a}}$ and $\widetilde{\mathbb{H-a}}$ in general, but the observations above indicates that their $S_{\tilde{\mathbb{a}}}$ and $S_{\widetilde{\mathbb{H-a}}}$ are the same as $S_{\mathbb{a}}$, which leads to \eqref{acpartw}.
Also, the non-abelian flat connections $\tilde{\beta}$ that are related by the Weyl action might be attached to the elements of the Weyl orbit of $t$, but it is unlikely.
If that were the case, for each poles there would be always two non-abelian flat connections $\tilde{\beta}$ that are in the same Weyl orbit with their own lifts, but we see from \eqref{ratstk} that there is just one exponential factors for the poles $m \in [-\lfloor 2P(H-t)/H \rfloor, -\lfloor 2Pt/H \rfloor -1]$.
So in this context, the sum would be over $\bbbeta$ in \eqref{st0w} with \eqref{cabnab}.

We note that $Z_{\tilde{a}}$ itself is not gauge invariant, but if we choose a gauge transformation as a based gauge transformation as discussed in \cite{Gukov-Marino-Putrov}, $Z_{\tilde{a}}$ may be a quantity that can be considered in the middle of the calculation in the context of resurgence.
Meanwhile, the total sum of them on the RHS of \eqref{acpartw} for a given abelian flat connection is a gauge invariant quantity.

If the expressions \eqref{acpartw} and \eqref{st0w} are taken, \eqref{aibrs} at $t \neq 0$ would be expressed as $I_{\tilde{\mathbb{t}}} + I_{\widetilde{\mathbb{H-t}}}$ where
\begin{align}
I_{\tilde{\mathbb{t}}} &= e^{2\pi i K S_{\mathbb{t}}} \int_{\gamma} dy e^{-\frac{K}{2\pi i} \frac{H}{P}(y+2\pi i \frac{P}{H}t)^2} \frac{\prod_{j=1}^3 e^{y/P_j}-e^{-y/P_j}}{e^{y}-e^{-y}}	\\
I_{\widetilde{\mathbb{H-t}}} &= e^{2\pi i K S_{\mathbb{t}}} \int_{\gamma} dy e^{-\frac{K}{2\pi i} \frac{H}{P}(y+2\pi i \frac{P}{H}(H-t))^2} \frac{\prod_{j=1}^3 e^{y/P_j}-e^{-y/P_j}}{e^{y}-e^{-y}}
\end{align}
and the residue part of \eqref{nabcont} would be expressed as
\begin{equation}
\frac{1}{2} m_{\tilde{\mathbb{t}}}^{\bbbeta_t} I_{\bbbeta_t}(|K|,0) + \frac{1}{2} m_{\widetilde{\mathbb{H-t}}}^{\bbbeta_{H-t}} I_{\bbbeta_{H-t}}(|K|,0)	\label{nabw}
\end{equation}
where
\begin{align} 
\begin{split}
\frac{1}{2} m_{\tilde{\mathbb{t}}}^{\bbbeta_t} I_{\bbbeta_t} &= 8 \pi e^{2 \pi i K S_{\mathbb{t}}} e^{-\frac{\pi i}{2} K \frac{H}{P} \big( m + \frac{2Pt}{H}\big)^2} (-1)^m \prod_{j=1}^3 \sin \frac{\pi m}{P_j}	\\
\frac{1}{2} m_{\widetilde{\mathbb{H-t}}}^{\bbbeta_{H-t}} I_{\bbbeta_{H-t}} &= 8 \pi e^{2 \pi i K S_{\mathbb{t}}} e^{-\frac{\pi i}{2} K \frac{H}{P} \big( m + \frac{2P(H-t)}{H}\big)^2} (-1)^m \prod_{j=1}^3 \sin \frac{\pi m}{P_j}	
\end{split}	\label{sep-cont}
\end{align}
for a given pole $y=m \pi i$ and here $\bbbeta_t$ and $\bbbeta_{H-t}$ are lifts of $\beta$.	\\

With above discussion in mind, we calculate transseries parameter associated to $y=-n\pi i$ for an example with $(P_1,P_2,P_3)=(2,3,7)$ and $H=5$ discussed in section \ref{ssec:rsgh2}.
We chose lifts such that $S_{\mathbb{t}} = \frac{P}{H}t^2$.
For the latter consideration, we would have

\begin{empheq}[left={\hspace{-26.5mm}m_{(a_0,0)}^{\bbbeta} = \empheqlbrace}]{alignat*=2}
&\ \text{for } \bbbeta=(\alpha_1,-\frac{5}{168}n^2)	&&\left\{ 
\begin{array}{l l l l l }
+1	&\quad n = 1, 41, 55, 71	&\ \text{mod } 84	\\
-1	&\quad n = 13, 29, 43, 83	&\ \text{mod } 84	
\end{array}	\right.			\\
&\ \text{for } \bbbeta=(\alpha_2,-\frac{5}{168}n^2)	&&\left\{ 
\begin{array}{l l l l l }
+1	&\quad n = 5, 19, 23, 37	&\ \text{mod } 84	\\
-1	&\quad n = 47, 61, 65, 79	&\ \text{mod } 84	
\end{array}	\right.			\\
&\ \text{for } \bbbeta=(\alpha_3,-\frac{5}{168}n^2)	&&\left\{ 
\begin{array}{l l l l l }
+1	&\quad n = 11, 17, 25, 31	&\ \text{mod } 84	\\
-1	&\quad n = 53, 59, 67, 73	&\ \text{mod } 84	
\end{array}	\right.		
\end{empheq}

\begin{empheq}[left={m_{(a_1,\frac{42}{5})}^{\bbbeta} = \empheqlbrace}]{alignat*=2}
&\ \text{for } \bbbeta=(\alpha_1,-\frac{5}{168}n^2 + n)	&&\left\{ 
\begin{array}{l l l l l }
-1	&\quad n = 29, 43, 83	&\ \text{mod } 84	\\
+1	&\quad n = 41, 55, 71	&\ \text{mod } 84	\\
+1	&\quad n = 1			&\ \text{mod } 84	\text{ with } n > 84		\\
-1	&\quad n = 13			&\ \text{mod } 84	\text{ with } n > 84	
\end{array}	\right.			\\
&\ \text{for } \bbbeta=(\alpha_2,-\frac{5}{168}n^2 + n)	&&\left\{ 
\begin{array}{l l l l l }
+1	&\quad n = 19, 23, 37	&\ \text{mod } 84	\\
-1	&\quad n = 47, 61, 65, 79	&\ \text{mod } 84	\\
+1	&\quad n = 5			&\ \text{mod } 84	\text{ with } n > 84	
\end{array}	\right.			\\
&\ \text{for } \bbbeta=(\alpha_3,-\frac{5}{168}n^2 + n)	&&\left\{ 
\begin{array}{l l l l l }
+1	&\quad n = 17, 25, 31	&\ \text{mod } 84	\\
-1	&\quad n = 53, 59, 67, 73	&\ \text{mod } 84	\\
+1	&\quad n = 11			&\ \text{mod } 84	\text{ with } n > 84
\end{array}	\right.		
\end{empheq}

\begin{empheq}[left={m_{(a_2,\frac{168}{5})}^{\bbbeta} = \empheqlbrace}]{alignat*=2}
&\ \text{for } \bbbeta=(\alpha_1,-\frac{5}{168}n^2 + 2n)	&&\left\{ 
\begin{array}{l l l l l }
+1	&\quad n = 41,55, 71		&\ \text{mod } 84	\\
-1	&\quad n = 43, 83		&\ \text{mod } 84	\\
+1	&\quad n = 1			&\ \text{mod } 84	\text{ with } n > 84		\\
-1	&\quad n = 13, 29		&\ \text{mod } 84	\text{ with } n > 84	
\end{array}	\right.			\\
&\ \text{for } \bbbeta=(\alpha_2,-\frac{5}{168}n^2 + 2n)	&&\left\{ 
\begin{array}{l l l l l }
+1	&\quad n = 37			&\ \text{mod } 84	\\
-1	&\quad n = 47, 61, 65, 79	&\ \text{mod } 84	\\
+1	&\quad n = 5, 19, 23		&\ \text{mod } 84	\text{ with } n > 84	
\end{array}	\right.			\\
&\ \text{for } \bbbeta=(\alpha_3,-\frac{5}{168}n^2 + 2n)	&&\left\{ 
\begin{array}{l l l l l }
-1	&\quad n = 53, 59, 67, 73		&\ \text{mod } 84	\\
+1	&\quad n = 11, 17, 25, 31		&\ \text{mod } 84	\text{ with } n > 84
\end{array}	\right.		
\end{empheq}

\begin{empheq}[left={m_{(a_3,\frac{168}{5})}^{\bbbeta} = \empheqlbrace}]{alignat*=2}
&\ \text{for } \bbbeta=(\alpha_1,-\frac{5}{168}n^2 + 3n - 84)	&&\left\{ 
\begin{array}{l l l l l }
+1	&\quad n = 55, 71			&\ \text{mod } 84	\\
-1	&\quad n = 83				&\ \text{mod } 84	\\
+1	&\quad n = 1, 41			&\ \text{mod } 84	\text{ with } n > 84		\\
-1	&\quad n = 13, 29, 43		&\ \text{mod } 84	\text{ with } n > 84	
\end{array}	\right.			\\
&\ \text{for } \bbbeta=(\alpha_2,-\frac{5}{168}n^2 + 3n - 84)	&&\left\{ 
\begin{array}{l l l l l }
-1	&\quad n = 61, 65, 79		&\ \text{mod } 84	\\
+1	&\quad n = 5, 19, 23, 37		&\ \text{mod } 84	\text{ with } n > 84	\\
-1	&\quad n = 47				&\ \text{mod } 84	\text{ with } n > 84
\end{array}	\right.			\\
&\ \text{for } \bbbeta=(\alpha_3,-\frac{5}{168}n^2 + 3n - 84)	&&\left\{ 
\begin{array}{l l l l l }
-1	&\quad n = 53, 59, 67, 73		&\ \text{mod } 84	\\
+1	&\quad n = 11, 17, 25, 31		&\ \text{mod } 84	\text{ with } n > 84
\end{array}	\right.		
\end{empheq}

\begin{empheq}[left={\hspace{1mm}m_{(a_4,\frac{42}{5})}^{\bbbeta} = \empheqlbrace}]{alignat*=2}
&\ \text{for } \bbbeta=(\alpha_1,-\frac{5}{168}n^2 + 4n + 126)	&&\left\{ 
\begin{array}{l l l l l }
+1	&\quad n = 71			&\ \text{mod } 84	\\
-1	&\quad n = 83			&\ \text{mod } 84	\\
+1	&\quad n = 1, 41, 55		&\ \text{mod } 84	\text{ with } n > 84		\\
-1	&\quad n = 13,29,43		&\ \text{mod } 84	\text{ with } n > 84	
\end{array}	\right.			\\
&\ \text{for } \bbbeta=(\alpha_2,-\frac{5}{168}n^2 + 4n + 126)	&&\left\{ 
\begin{array}{l l l l l }
-1	&\quad n = 79			&\ \text{mod } 84	\\
+1	&\quad n = 5, 19, 23, 37	&\ \text{mod } 84	\text{ with } n > 84	\\
-1	&\quad n = 47, 61, 65	&\ \text{mod } 84	\text{ with } n > 84
\end{array}	\right.			\\
&\ \text{for } \bbbeta=(\alpha_3,-\frac{5}{168}n^2 + 4n + 126)	&&\left\{ 
\begin{array}{l l l l l }
-1	&\quad n = 73			&\ \text{mod } 84	\\
+1	&\quad n = 11, 17, 25, 31	&\ \text{mod } 84	\text{ with } n > 84	\\
-1	&\quad n = 53, 59, 67	&\ \text{mod } 84	\text{ with } n > 84
\end{array}	\right.		
\end{empheq}
For the former consideration, transseries parameters are given by $m_{(a_0,0)}^{\bbbeta}$, $m_{(a_1,\frac{42}{5})}^{\bbbeta}+m_{(a_4,\frac{42}{5})}^{\bbbeta}$, and $m_{(a_2,\frac{168}{5})}^{\bbbeta}+m_{(a_3,\frac{168}{5})}^{\bbbeta}$ for $t=0,1$, and $2$, respectively.


\subsection{Transseries parameters for a rational level $K$}
\label{app:trratk}

We also calculate transseries parameter for the case of the Seifert rational homology spheres when the level $K$ is taken to be a rational number.
The contributions from non-abelian flat connections and transseries parameters for the case of $H \geq 2$ can be calculated from \eqref{irsr-res} for odd $H$, 
\begin{align}
\begin{split}
&\hspace{-18mm} 8 \pi \sum_{v=0}^{s-1} \Bigg( e^{2 \pi i \frac{r}{s} \frac{P}{H}(Hv)^2} \sum_{m=-\infty}^{0} e^{-\frac{\pi i}{2} \frac{r}{s} \frac{H}{P} m^2} (-1)^m \prod_{j=1}^3 \sin \frac{\pi m}{P_j}	\\
&\hspace{-8mm}+\sum_{u=1}^{\frac{H-1}{2}} \bigg( \sum_{m=-\infty}^{-\lfloor \frac{2Pu}{H} \rfloor -1} e^{-\frac{\pi i}{2} \frac{r}{s} \frac{H}{P} \big( m + \frac{2Pu}{H}\big)^2} (-1)^m \prod_{j=1}^3 \sin \frac{\pi m}{P_j}	+ \sum_{m=-\infty}^{-\lfloor \frac{2P(H-u)}{H} \rfloor -1} e^{-\frac{\pi i}{2} \frac{r}{s} \frac{H}{P} \big( m + \frac{2P(H-u)}{H}\big)^2} (-1)^m \prod_{j=1}^3 \sin \frac{\pi m}{P_j} \bigg)
\Bigg)	\,	.
\end{split}	\label{irsr-reso}
\end{align}
When $H$ is even, we have
\begin{align}
\begin{split}
&\hspace{-10mm} 8 \pi \sum_{v=0}^{s-1} \Bigg( e^{2 \pi i \frac{r}{s} \frac{P}{H}(Hv)^2} \sum_{m=-\infty}^{0} e^{-\frac{\pi i}{2} \frac{r}{s} \frac{H}{P} m^2} (-1)^m \prod_{j=1}^3 \sin \frac{\pi m}{P_j}	+ e^{2 \pi i \frac{r}{s} HP (v+\frac{1}{2})^2} \sum_{m=-\infty}^{-P -1} e^{-\frac{\pi i}{2} \frac{r}{s} \frac{H}{P} ( m + P )^2} (-1)^m \prod_{j=1}^3 \sin \frac{\pi m}{P_j}	\\
&\hspace{0mm} +\sum_{u=1}^{\frac{H}{2}-1} e^{2 \pi i \frac{r}{s} \frac{P}{H} (Hv+u)^2} \bigg( \sum_{m=-\infty}^{-\lfloor \frac{2Pu}{H} \rfloor -1} e^{-\frac{\pi i}{2} \frac{r}{s} \frac{H}{P} \big( m + \frac{2Pu}{H}\big)^2} (-1)^m \prod_{j=1}^3 \sin \frac{\pi m}{P_j}	\\
&\hspace{40mm}	+ \sum_{m=-\infty}^{-\lfloor \frac{2P(H-u)}{H} \rfloor-1} e^{-\frac{\pi i}{2} \frac{r}{s} \frac{H}{P} \big( m + \frac{2P(H-u)}{H}\big)^2} (-1)^m \prod_{j=1}^3 \sin \frac{\pi m}{P_j} \bigg)
\Bigg)	\,	.
\end{split}	\label{irsr-rese}
\end{align}
After the regularization, \eqref{irsr-reso} and \eqref{irsr-rese} become, respectively,
\begin{align}
\begin{split}
&-4 \pi \sum_{v=0}^{s-1} e^{2\pi i \frac{r}{s} P H v^2 } \sum_{n=0}^{2Ps} e^{-\frac{\pi i}{2} \frac{r}{s} \frac{H}{P} n^2} (-1)^n \Big( 1 - \frac{n}{Ps} \Big) \prod_{j=1}^3 \sin \frac{\pi n}{P_j}	\\
&-8 \pi \sum_{v=0}^{s-1} \sum_{u=1}^{\frac{H-1}{2}} e^{2\pi i \frac{r}{s} \frac{P}{H} (Hv+u)^2 }
\bigg[
\frac{1}{2} \sum_{n=0}^{2Ps} \Big( e^{-\frac{\pi i}{2} \frac{r}{s} \frac{H}{P} \big( -n+\frac{2Pu}{H} \big)^2} + e^{-\frac{\pi i}{2} \frac{r}{s} \frac{H}{P} \big( n+\frac{2Pu}{H} \big)^2} \Big) (-1)^n \Big( 1-\frac{n}{Ps} \Big) \prod_{j=1}^3 \sin \frac{\pi n}{P_j}	\\
&\hspace{50mm}-\sum_{n=0}^{\lfloor \frac{2Pu}{H} \rfloor} e^{-\frac{\pi i}{2} \frac{r}{s} \frac{H}{P} \big( -n+\frac{2Pu}{H} \big)^2} (-1)^n \prod_{j=1}^3 \sin \frac{\pi n}{P_j}	\\
&\hspace{50mm}-\sum_{n=0}^{\lceil \frac{2Pu}{H} \rceil-1} e^{-\frac{\pi i}{2} \frac{r}{s} \frac{H}{P} \big( -n+\frac{2Pu}{H} \big)^2} (-1)^n \prod_{j=1}^3 \sin \frac{\pi n}{P_j}
\bigg]	\label{resh2odd}
\end{split}
\end{align}
and 
\begin{align}
\begin{split}
&-4 \pi \sum_{v=0}^{s-1} e^{2\pi i \frac{r}{s} P H v^2 } \sum_{n=0}^{2Ps} e^{-\frac{\pi i}{2} \frac{r}{s} \frac{H}{P} n^2} (-1)^n \Big( 1 - \frac{n}{Ps} \Big) \prod_{j=1}^3 \sin \frac{\pi n}{P_j}	\\
&-8 \pi \sum_{v=0}^{s-1} \sum_{u=1}^{\frac{H}{2}-1} e^{2\pi i \frac{r}{s} \frac{P}{H} (Hv+u)^2 }
\bigg[
\frac{1}{2} \sum_{n=0}^{2Ps} \Big( e^{-\frac{\pi i}{2} \frac{r}{s} \frac{H}{P} \big( -n+\frac{2Pu}{H} \big)^2} + e^{-\frac{\pi i}{2} \frac{r}{s} \frac{H}{P} \big( n+\frac{2Pu}{H} \big)^2} \Big) (-1)^n \Big( 1-\frac{n}{Ps} \Big) \prod_{j=1}^3 \sin \frac{\pi n}{P_j}	\\
&\hspace{50mm}-\sum_{n=0}^{\lfloor \frac{2Pu}{H} \rfloor} e^{-\frac{\pi i}{2} \frac{r}{s} \frac{H}{P} \big( -n+\frac{2Pu}{H} \big)^2} (-1)^n \prod_{j=1}^3 \sin \frac{\pi n}{P_j}	\\
&\hspace{50mm}-\sum_{n=0}^{\lceil \frac{2Pu}{H} \rceil-1} e^{-\frac{\pi i}{2} \frac{r}{s} \frac{H}{P} \big( -n+\frac{2Pu}{H} \big)^2} (-1)^n \prod_{j=1}^3 \sin \frac{\pi n}{P_j}
\bigg]	\\
&-4 \pi \sum_{v=0}^{s-1} e^{2\pi i \frac{r}{s} P H \big(v+\frac{1}{2}\big)^2 } \sum_{n=0}^{2Ps} e^{-\frac{\pi i}{2} \frac{r}{s} \frac{H}{P} n^2} (-1)^n \Big( 1 - \frac{n}{Ps} \Big) \prod_{j=1}^3 \sin \frac{\pi n}{P_j}	\,	.	\label{resh2even}
\end{split}
\end{align}
We take an example with $(P_1,P_2,P_3)=(2,3,7)$ with $H=5$ and for concreteness we choose $s=11$.

We first sort poles, $y=-n \pi i$, that give the same exponent of $e^{-2 \pi i \frac{1}{4} \frac{H}{P} \frac{r}{s} n^2}$, which may be grouped in three larger sets of poles that give the same $-\frac{1}{4} \frac{H}{P} n^2$ mod $1$,
\begin{empheq}[left = {\hspace{-8mm} }]{alignat=3}
& \alpha_1	\rightarrow
&& 
\left\{
\begin{array}{l l l}
\alpha_{1,1} :	&	1,43,155,197,265,307,419,461,463,505,617,659,727,769,881,923,		\\
\alpha_{1,2} :	&	13,97,167,211,251,295,365,449,475,559,629,673,713,757,827,911,	\\
\alpha_{1,3} :	&	29,125,139,169,293,323,337,433,491,587,601,631,755,785,799,895,	\\
\alpha_{1,4} :	&	41,85,113,223,239,349,377,421,503,547,575,685,701,811,839,883,	\\
\alpha_{1,5} :	&	55,209,253,407,517,671,715,869,	\\
\alpha_{1,6} :	&	71,83,127,181,281,335,379,391,533,545,589,643,743,797,841,853
\end{array}
\right.	\label{pr2371}
\\
&
\alpha_2	\rightarrow
&&
\left\{
\begin{array}{l l}
\alpha_{2,1} :	&	5,61,149,215,247,313,401,457,467,523,611,677,709,775,863,919,	\\
\alpha_{2,2} :	&	19,47,107,173,289,355,415,443,481,509,569,635,751,817,877,905,	\\
\alpha_{2,3} :	&	23,65,89,131,331,373,397,439,485,527,551,593,793,835,859,901,	\\
\alpha_{2,4} :	&	37,103,191,205,257,271,359,425,499,565,653,667,719,733,821,887,	\\
\alpha_{2,5} :	&	79,145,163,229,233,299,317,383,541,607,625,691,695,761,779,845,	\\
\alpha_{2,6} :	&	121,187,275,341,583,649,737,803
\end{array}
\right.	\label{pr2372}
\\
&
\alpha_3	\rightarrow
&&
\left\{
\begin{array}{l l}
\alpha_{3,1} :	&	11,143,319,451,473,605,781,913,	\\
\alpha_{3,2} :	&	17,115,137,193,269,325,347,445,479,577,599,655,731,787,809,907,	\\
\alpha_{3,3} :	&	25,151,157,179,283,305,311,437,487,613,619,641,745,767,773,899,	\\
\alpha_{3,4} :	&	31,53,101,185,277,361,409,431,493,515,563,647,739,823,871,893,	\\
\alpha_{3,5} :	&	59,73,95,227,235,367,389,403,521,535,557,689,697,829,851,865,	\\
\alpha_{3,6} :	&	67,109,199,221,241,263,353,395,529,571,661,683,703,725,815,857
\end{array}
\right.	\label{pr2373}
\end{empheq}
mod 924.

The contributions from non-abelian flat connections attached to abelian flat connections labelled by Weyl orbits of $u$, which we denote as $a$, are calculated from \eqref{resh2odd}.
Denoting $\zeta=\zeta_{11} = e^{\frac{2\pi i}{11}}$, the sum of residues for each set of poles $(\alpha_{j,l})_{l=1, \ldots, 6}$, $j=1,2,3$ above are given by
\begin{empheq}[left = {\hspace{-38.0mm}\text{for } a=0 \ \empheqlbrace} ]{alignat*=3} 
(\alpha_{1,l})_{l}	\rightarrow
& \
(0,-2 \zeta ^{6 r},\zeta ^{8 r},\zeta ^{5 r},\zeta ^{9 r},-\zeta ^{4 r})
\\
(\alpha_{2,l})_{l}	\rightarrow
& \
(-1,\zeta ^r,0,2 \zeta ^{4 r},-2 \zeta ^{2 r},\zeta ^{5 r})
\\
(\alpha_{3,l})_{l}	\rightarrow
& \
(0,\zeta ^{6 r},0,\zeta ^{8 r},-\zeta ^{10 r},0)
\end{empheq}
\begin{empheq}[left = \hspace{0mm}{ \text{for } a=1 \ \empheqlbrace}]{alignat*=3} 
(\alpha_{1,l})_{l}	\rightarrow
& \
(-\zeta ^{8 r},-2-\zeta ^{4 r},1+\zeta ^{8 r},\zeta ^{4 r}+\zeta ^{9 r},2 \zeta ^{5 r},-\zeta ^{5 r}-\zeta ^{6 r})
   \\
(\alpha_{2,l})_{l}	\rightarrow
& \
(-2 \zeta ^r-\zeta ^{2 r},1+\zeta ^{5 r},0,2 \zeta ^{4 r}+2 \zeta ^{7 r},-2-2 \zeta ^{7 r},2 \zeta ^r)
\\
(\alpha_{3,l})_{l}	\rightarrow
& \
(-\zeta ^{7 r},\zeta ^{7 r}+\zeta ^{8 r},0,\zeta ^{2 r}+\zeta ^{6 r},-\zeta ^{2 r}-\zeta^{10 r},0)
\end{empheq}
\begin{empheq}[left = {\hspace{4.5mm} \text{for } a=2 \ \empheqlbrace }]{alignat*=3} 
(\alpha_{1,l})_{l}	\rightarrow
& \
(-\zeta ^{8 r},-\zeta ^{5 r}-2 \zeta ^{8 r},1+2 \zeta ^{6 r},\zeta ^{5 r}+\zeta ^{6 r},2\zeta ^{4 r},-1-\zeta ^{9 r})
\\
(\alpha_{2,l})_{l}	\rightarrow
& \
(-2 \zeta ^{5 r}-\zeta ^{7 r},\zeta ^{2 r},-\zeta ^{4 r},2 \zeta ^{2 r}+2 \zeta ^{7 r},-2\zeta ^r-2 \zeta ^{4 r},2)
\\
(\alpha_{3,l})_{l}	\rightarrow
& \
(-\zeta ^{6 r},1,-\zeta ^{8 r},\zeta ^{7 r},-\zeta ^{2 r}-\zeta ^{8 r},0)
\end{empheq}
where for each $\alpha_{j,l}$ overall factors $Z'_{\alpha_j}$
\begin{align}
Z'_{\alpha_1} &= 8 \sqrt{3} \pi \sin \left(\frac{\pi }{7}\right) e^{-2 \pi  i \frac{1}{4} \frac{r}{11} \frac{5}{42}  1^2}g(210 r,11)	\,	,	\label{237s11ovft1}	\\
Z'_{\alpha_2} &= -8 \sqrt{3} \pi \cos \left(\frac{3 \pi }{14}\right) e^{-2 \pi  i \frac{1}{4} \frac{r}{11} \frac{5}{42}5^2}g(210 r,11)	\,	,	\label{237s11ovft2}\\
Z'_{\alpha_3} &=-8 \sqrt{3} \pi  \cos \left(\frac{\pi }{14}\right) e^{-2 \pi  i \frac{1}{4} \frac{r}{11} \frac{5}{42}11^2}g(210 r,11)	\,	.	\label{237s11ovft3}
\end{align}
are multiplied.
The sum of residues for $\alpha_{j,l}$, $l=1, \ldots, 6$ for a given $j$, $j=1,2,3$ is given by
\begin{align}
&(-1+\zeta ^{4 r}+2 \zeta ^{5 r}-2 \zeta ^{8 r}+\zeta ^{9 r}) \, Z'_{\alpha_1}	\label{tot-nonab-rat-237-1}\\
&(-\zeta ^r+\zeta ^{4 r}+\zeta ^{7 r}) \, Z'_{\alpha_2}	\\
&(1-\zeta ^{2 r}+\zeta ^{6 r}+\zeta ^{7 r}-2 \zeta ^{10 r}) \, Z'_{\alpha_3} 	\,	,
\end{align}
respectively.
When $s$ is taken to be 1 above, they agree with the results for the integer $K$ discussed in section \ref{ssec:rsgh2}.	\\

We can calculate the transseries parameter associated to the pole $y=-n \pi i$ from \eqref{irsr-reso} or \eqref{irsr-rese}.
Given \eqref{pr2371}, \eqref{pr2372}, and \eqref{pr2373}, the transseries parameter for $u=0$ is given by
\begin{empheq}[left = {\hspace{-8mm} }]{alignat=3}
& \alpha_1	\rightarrow
&& 
\left\{
\begin{array}{l l l}
\alpha_{1,1}	&\leadsto	&1,-1,1,-1,-1,1,-1,1,-1,1,-1,1,1,-1,1,-1	\\
\alpha_{1,2}	&\leadsto	&-1,-1,-1,-1,-1,-1,-1,-1,1,1,1,1,1,1,1,1	\\
\alpha_{1,3}	&\leadsto	&-1,1,1,1,1,1,1,-1,1,-1,-1,-1,-1,-1,-1,1	\\
\alpha_{1,4}	&\leadsto	&1,1,-1,1,1,-1,1,1,-1,-1,1,-1,-1,1,-1,-1	\\
\alpha_{1,5}	&\leadsto	&1,1,1,1,-1,-1,-1,-1	\\
\alpha_{1,6}	&\leadsto	&1,-1,-1,-1,-1,-1,-1,1,-1,1,1,1,1,1,1,-1
\end{array}
\right.	\label{pr2371m}
\\
&
\alpha_2	\rightarrow
&&
\left\{
\begin{array}{l l l}
\alpha_{2,1}	&\leadsto	&1,-1,-1,-1,-1,-1,-1,1,-1,1,1,1,1,1,1,-1	\\
\alpha_{2,2}	&\leadsto	&1,-1,1,1,1,1,-1,1,-1,1,-1,-1,-1,-1,1,-1	\\
\alpha_{2,3}	&\leadsto	&1,-1,1,-1,-1,1,-1,1,-1,1,-1,1,1,-1,1,-1	\\
\alpha_{2,4}	&\leadsto	&1,1,1,1,1,1,1,1,-1,-1,-1,-1,-1,-1,-1,-1	\\
\alpha_{2,5}	&\leadsto	&-1,-1,-1,-1,-1,-1,-1,-1,1,1,1,1,1,1,1,1	\\
\alpha_{2,6}	&\leadsto	&1,1,1,1,-1,-1,-1,-1
\end{array}
\right.	\label{pr2372m}
\\
&
\alpha_3	\rightarrow
&&
\left\{
\begin{array}{l l l}
\alpha_{3,1}	&\leadsto	&1,-1,-1,1,-1,1,1,-1	\\
\alpha_{3,2}	&\leadsto	&1,1,-1,1,1,-1,1,1,-1,-1,1,-1,-1,1,-1,-1	\\
\alpha_{3,3}	&\leadsto	&1,-1,-1,1,1,-1,-1,1,-1,1,1,-1,-1,1,1,-1	\\
\alpha_{3,4}	&\leadsto	&1,-1,1,1,1,1,-1,1,-1,1,-1,-1,-1,-1,1,-1	\\
\alpha_{3,5}	&\leadsto	&-1,-1,1,-1,-1,1,-1,-1,1,1,-1,1,1,-1,1,1	\\
\alpha_{3,6}	&\leadsto	&-1,1,1,-1,-1,1,1,-1,1,-1,-1,1,1,-1,-1,1
\end{array}
\right.	\label{pr2373m}
\end{empheq}
where $Z_{\alpha_{1,l}} = 8 \sqrt{3} \pi  \sin \frac{\pi }{7}$, $Z_{\alpha_{2,l}} = -8 \sqrt{3} \pi  \cos \frac{3 \pi }{14}$, and $Z_{\alpha_{2,l}} = -8 \sqrt{3} \pi  \cos \frac{\pi }{14}$, $l=1, \ldots, 6$.
Also, a lift $\bbbeta$ of $\beta$ is read off from the exponent $PH v^2 -\frac{1}{4} \frac{H}{P}m^2$ of the exponential factors $\sum_{v=0}^{s-1} e^{2 \pi i \frac{r}{s} \frac{P}{H}(Hv)^2} e^{-\frac{\pi i}{2} \frac{r}{s} \frac{H}{P} m^2}$.

For other $u$'s, some poles in \eqref{pr2371}, \eqref{pr2372}, and \eqref{pr2373} are not included, which are 
\begin{equation}
\begin{tabular}{l l}
$u=1$	&$1,5,11,13$	\\
$u=2$	&$1, 5, 11, 13, 17, 19, 23, 25, 29, 31$	\\
$u=3$	&$1, 5, 11, 13, 17, 19, 23, 25, 29, 31, 37, 41, 43, 47$	\\
$u=4$	&$1, 5, 11, 13, 17, 19, 23, 25, 29, 31, 37, 41, 43, 47, 53, 55, 59, 61, 65, 67$
\end{tabular}	\label{exclp}
\end{equation}
so corresponding transseries parameters are zero.
Transseries parameters associated to \eqref{exclp} mod 924 with $m > 924$ and to other poles can be read off from \eqref{pr2371m}, \eqref{pr2372m}, and \eqref{pr2373m}.
The lift is read off from $\frac{P}{H}(Hv+u)^2 - \frac{1}{4}\frac{H}{P} (-m+\frac{2Pu}{H})^2$ of $\sum_{v=0}^{s-1} e^{2 \pi i \frac{r}{s} \frac{P}{H}(Hv+u)^2} e^{-\frac{\pi i}{2} \frac{r}{s} \frac{H}{P} (-m+\frac{2Pu}{H})^2}$, $u=1,2,3$, and $4$.\footnote{For reference, $\sum_{v=0}^{s-1} e^{2 \pi i \frac{r}{s} \frac{P}{H}(Hv+u)^2} = \sum_{v=0}^{s-1} e^{2 \pi i \frac{r}{s} \frac{P}{H}(Hv+H-u)^2}$.}

\end{appendices}

\bibliographystyle{JHEP}
\bibliography{ref}

\end{document}